\documentstyle[11pt]{article}
\input epsf

\begin{document}
\def\bsigma{\mbox{\boldmath $\sigma$}} 
\def\balpha{{\mbox{\boldmath $\alpha$}}}
\def\bbeta{{\mbox{\boldmath $\beta$}}}
\def\bgamma{{\mbox{\boldmath $\gamma$}}}
\def\la{\mathrel{\mathpalette\fun <}}
\def\ga{\mathrel{\mathpalette\fun >}}
\def\fun#1#2{\lower3.6pt\vbox{\baselineskip0pt\lineskip.9pt
  \ialign{$\mathsurround=0pt#1\hfil##\hfil$\crcr#2\crcr\sim\crcr}}}

\baselineskip 20pt
\rightline{CU-TP-903}
\rightline{hep-th/9908097}
\vskip 2cm

\makeatletter
\renewcommand\theequation{\thesection.\arabic{equation}}
\@addtoreset{equation}{section}
\makeatother

\centerline{{\Large\bf Massive Monopoles and Massless Monopole
Clouds}\footnote{Talks delivered at the International Workshop on
Mathematical and Physical Aspects of Nonlinear Field Theories, Seoul,
Korea, February 1998}}
\bigskip
\centerline{ Erick J. Weinberg}
\medskip
\centerline{ Department of Physics, Columbia University}
\centerline{ New York, NY 10027, USA}
 
\bigskip
\centerline{\bf Abstract}
\medskip
\centerline{\vbox{\hsize=5truein 
\baselineskip=12truept \tenrm \noindent 
Magnetic monopole solutions naturally arise in the context of
spontaneously broken gauge theories.  When the unbroken symmetry
includes a non-Abelian subgroup, investigation of the low-energy
monopole dynamics by means of the moduli space approximation reveals
degrees of freedom that can be attributed to massless monopoles.
These do not correspond to distinct solitons, but instead are
manifested as a cloud of non-Abelian field surrounding one or more
massive monopoles.  In these talks I explain how one is led to these
solutions and then describe them in some detail.}}

\baselineskip 20pt

\section{Introduction}
  
One-particle states arise in the spectra of weakly coupled
quantum field theories in two rather different ways.  By quantizing
the small oscillations about the vacuum, one finds the states, with a
characteristic mass $m$, that correspond to the quanta of
the fundamental fields of the theory.  It may also happen that the
classical field equations of the theory possess localized solutions with
energies of order $m/\lambda$, where $\lambda$ is a typical small coupling of
the theory; these soliton solutions also give rise to particles in the quantum
theory.  At first sight, these two classes of particles appear quite
different: the former seem to be point particles with no internal structure,
while the latter are extended objects described by a classical field profile
$\phi({\bf r})$.

However, these distinctions are not quite so clearcut.  On the one
hand, in an interacting theory even the fundamental point particles
can be viewed as having a partonic substructure that evolves with
momentum scale according to the DGLAP equations.  On the other, one
can analyze the behavior of the soliton states in terms of the normal modes
of small fluctuations about the soliton.  The modes in the continuum
part of the spectra can be interpreted as scattering states of
elementary quanta in the presence of the soliton; there may also be
discrete eigenvalues corresponding to quanta bound to the soliton.
This leaves only a small number of zero frequency modes whose
quantization entails the introduction of the collective coordinates that
may be viewed as the fundamental degrees of freedom of the
soliton.

These considerations suggest that the particle states built from
solitons and those based the elementary quanta do not differ in any
essential way.  Indeed, it can happen that states that appear as
elementary quanta in one formulation of the theory correspond to
solitons in another.  The classic example of this is the
correspondence between the sine-Gordon model and the Thirring model
\cite{sineG}.  Of more immediate relevance to my talks is the
conjecture by Montonen and Olive \cite{dual} that certain theories
possess an exact electromagnetic duality relating 
magnetically charged solitons and electrically charged elementary
quanta.

If there is such a duality, then one would expect the classical
solutions to display particle-like properties.  In
particular, one would expect the classical solutions with higher
charges to have a structure consistent with an interpretation
in terms of component solitons of minimal charge.  This is indeed
found to be the case in many theories.  However, in the course of
studying magnetic monopoles in the context of larger gauge groups
\cite{klee,klee2}, Kimyeong Lee, Piljin Yi, and I found
\cite{nonabelian} that in some theories there are classical solutions
that do not quite fit this picture.  As I will explain below, there is
a sense in which these solutions can be understood as multimonopole solutions
containing both 
massive and massless magnetic monopoles.  While the massive components
are quite evident when one examines the classical solutions, the
massless monopoles appear to lose their individual identity and merge
into a ``cloud'' of non-Abelian fields.  Because these massless
monopoles can be viewed as the duals to the massless gauge bosons of
the theory, a better understanding of the nature of these unusual
solutions may well provide deeper insight into the properties of
non-Abelian gauge theories.

In these talks I will explain how one is led to these solutions and
then describe them in some detail.  I begin in the next section by
reviewing some general properties of magnetic monopoles.  In Sec.~3, I
describe the Bogomolny-Prasad-Sommerfield, or BPS, limit
\cite{bps} and its application to multimonopole solutions in an
$SU(2)$ gauge theory.  The extension of these results to larger gauge
groups is discussed in Sec.~4.  The treatment of low-energy monopole
dynamics by means of the moduli space approximation is discussed in
Sec.~5, while the actual determination of some moduli space metrics is
described in Sec.~6.  The theories where one actually encounters
evidence of massless monopoles are those in which the unbroken gauge
symmetry has a non-Abelian component.  These are discussed in Sec.~7.
Explicit examples of solutions in which the massless monopoles appear
to condense into a non-Abelian cloud are described in Secs.~8 and 9.
Section~10 contains some concluding remarks.

\section{Magnetic monopoles}

     In the absence of sources, Maxwell's equations display a symmetry
under the interchange of electric and magnetic fields.  This suggests
that there might also be a symmetry in sources, and that in addition
to the familiar electric charges there might also be magnetically
charged objects, usually termed magnetic monopoles, that act as
sources for magnetic fields.  A static monopole with magnetic charge
$Q_M$ would give rise to a Coulomb magnetic field
\begin{equation}
    B_i = {Q_M\over 4\pi} { {\hat r}_i \over r^2}  \, .
\label{bfield}
\end{equation}
 
     In the canonical treatment of the behavior of charged particles
in a magnetic field, either classically or quantum mechanically, it is
most convenient to express the magnetic field as the curl of a vector
potential ${\cal A}_i$.  For the magnetic field of Eq.~(\ref{bfield}), a
suitable choice is
\begin{equation}
    {\cal A}_i = -\epsilon_{ij3}{Q_M\over 4\pi} { {\hat r}_j \over r} {(1 -\cos
             \theta)  \over \sin\theta} \, .
\end{equation}
Note that this is singular along the negative $z$-axis.  This ``Dirac
string'' singularity is an inevitable consequence of trying to express a field
with nonvanishing divergence as the curl of a potential; any
potential leading to Eq.~(\ref{bfield}) will have a similar
singularity along some curve running from the position of the monopole
out to infinity.  Physically, this singularity is a difficulty only
if it actually observable.  In classical physics, where only the magnetic
field, and not the vector potential, is measurable, it causes no problem.
However, there are quantum mechanical interference effects that are sensitive to
the quantity \begin{equation}
   U= \exp\left[{ie } \oint_C {\cal A}_i dl_i \right]
\end{equation}
where $e$ is the electric charge of some particle and the integration
is around any closed curve.  If $C$ is taken to be an infinitesimal
closed curve in a region where ${\cal A}_i$ is nonsingular, $U$ is clearly
equal to unity.  On the other hand, if the integral is taken around an
infinitesimal closed curve encircling the Dirac string, $U$ is not
equal to unity, and the string is thus observable, unless the magnetic
charge obeys the Dirac quantization condition\footnote{I am assuming units
in which $\hbar=1$; otherwise there is a additional factor of $\hbar$ on the
right hand side.}
\begin{equation}
   Q_M = {4 \pi \over e} \left( {n\over 2}\right)
\label{diraccondition}
\end{equation}
for some integer $n$.  If we want the string to be unobservable, this
condition must hold for all possible electric charges.  This is only 
possible if all electric charges are integer multiples of some minimum
charge for which Eq.~(\ref{diraccondition}) is satisfied.  Thus, the
existence of a single monopole in the universe would be sufficient to
explain the observed quantization of electric charge.

     There is an alternative approach that avoids the appearance of
string singularities \cite{wuyang}. Instead, one introduces two gauge patches,
one excluding the negative $z$-axis and one excluding the positive
$z$-axis, and in each one chooses a vector potential that is
nonsingular in that region.  In the overlap of the two regions, the
two vector potentials can differ only by the gauge transformation that
relates the two patches.  In order that this gauge transformation be
single-valued in the overlap region, Eq.~(\ref{diraccondition}) must
hold.

     One can always incorporate magnetic monopoles into a theory with
electrically charged particles simply by postulating a new species of
fundamental particles.  However, it turns out \cite{hooft} that monopoles are
already implicit in many theories with electrically charged
fundamental fields.  In these theories, the classical field equations
have localized finite energy solutions with magnetic charge that
correspond to one-particle states of the quantized theory.  Although 
topological arguments are usually used to demonstrate the existence of these
solutions, their existence and many of their features can in fact be understood
on the basis of energetic arguments alone \cite{LeeEW}.
 
     To begin, note that the Coulomb magnetic field Eq.~(\ref{bfield}) has a
$1/r^2$ singularity at the origin.  In contrast to the Dirac string, this a
true physical singularity, as can be seen by noting that it leads to a
$1/r^4$ divergence in the energy density ${\cal E}= {1\over 2}{\bf
B}^2$.  This singularity must somehow be tamed if finite energy
classical solutions are to exist.  One approach might be to replace
the point magnetic charge by a charge distribution, but the Dirac
quantization condition forbids such continuous charge distributions in
theories with both electric and magnetic charges.

However, there is another possibility for removing the divergence.
When placed in a magnetic field, a magnetic dipole ${\bf d}$ acquires
an energy $-{\bf d\cdot B}$.  Thus, the singular energy
density in the magnetic field might be cancelled by introducing a
suitable (singular) distribution of magnetic dipoles.  This idea can
be implemented by introducing a complex vector field $\bf W$ with
electric charge $e$ and a magnetic dipole density ${\bf d}=ieg {\bf
W}^* \times {\bf W}$, with $g$ a real constant that for the moment can
be taken to be arbitrary.  Since we want there to be a lower bound on
the energy, the energy density must also contain terms of higher order
in $\bf d$.  In particular, adding a term ${\bf d}^2/2$ allows the
$1/r^4$ divergence of the energy density to be cancelled if $|{\bf W}|
\sim 1/r$ near the origin.

Since we want the $W$ field to be localized within a finite region, the
energy density should contain a mass term of the form $M_W^2 |{\bf
W}|^2$.  However, this would give a $1/r^2$ contribution to $\cal E$ near the
origin.  This singularity can be eliminated by allowing the $W$ mass
to be dependent on some spatially varying field $\phi$.  In
particular, let us assume that $M_W=G\phi$, where $G$ is a constant
and the scalar field potential $V(\phi)$ is minimized by $\phi=v\ne
0$.  Finiteness of the energy then implies that at large distances
$\phi \approx v$ and $M_W\ne 0$, but at $r=0$ both $\phi$ and $M_W$
can be taken to vanish.  Provided that the contribution from the
gradients of the fields introduce no additional singularities (which can be
arranged), the energy density will then be nonsingular everywhere.

An energy density of the sort described here can be obtained from a Lagrangian
density of the form 
\begin{equation}
    {\cal L}=  -{1\over 4} ({\cal F}_{\mu\nu} -iegW^*_\mu W_\nu)^2 
            -{1\over 2} |{\cal D}_\mu W_\nu -{\cal D}_\nu W_\mu |^2  
            + G^2 \phi^2 |W_\mu|^2   -{1\over 2} (\partial_\mu \phi)^2
          - V(\phi)  
\label{EMlag}
\end{equation}
where ${\cal D}_\mu = \partial_\mu +ie{\cal A}_\mu$ and ${\cal F}_{\mu\nu} =
\partial_\mu {\cal A}_\nu - \partial_\nu {\cal A}_\mu$ are the electromagnetic
covariant derivative and electromagnetic field strength.  
A solution of the resulting Euler-Lagrange equations
that carries unit
magnetic charge (i.e., $Q_M=4\pi/e$) can be obtained by introducing the ansatz 
\begin{eqnarray}
    {\cal A}_i &=& -\epsilon_{ij3}{ {\hat r}_j \over er} {(1 -\cos \theta)  \over
           \sin\theta}  \cr
    W_1 &=& -{i\over \sqrt{2}} {u(r) \over er} 
       \left[ 1 -e^{i\phi} \cos\phi (1-\cos\theta) \right]  \cr
    W_2 &=& {1\over \sqrt{2}} {u(r) \over er} 
       \left[ 1 +e^{i\phi} \sin\phi (1-\cos\theta)  \right] \cr
    W_3 &=& {i\over \sqrt{2}} {u(r) \over er} e^{i\phi} \sin\theta \cr
    \phi &=& h(r)  \, .
\end{eqnarray}
Substitution of this ansatz into the Euler-Lagrange equations leads to
a pair of coupled second order ordinary differential equations that
can be solved numerically to yield a finite energy solution.  
However, that the Dirac string singularity still remains.

A very important special case is obtained by setting $g=2$ and $G=e$. 
With this choice of parameters, the theory described by Eq~(\ref{EMlag}) is in
fact an $SU(2)$ gauge theory in disguise.  Let us define the components of an
$SU(2)$ gauge field $A_\mu \equiv A_\mu^a T^a$ and a triplet Higgs field $\Phi
\equiv \Phi^a T^a$ by  
\begin{eqnarray}
    A_\mu^1 + i A_\mu^2 &=& W_\mu,  \qquad A_\mu^3 = {\cal A}_\mu  \cr
    \Phi^a &=& \delta^{a3} \phi(r)
\label{stringgauge}
\end{eqnarray} 
where the $T^a$ are the generators of $SU(2)$.
The Lagrangian (\ref{EMlag}) can then be rewritten in the form
\begin{equation}
   {\cal L} = -{1\over 4} {\rm Tr}\,F_{\mu\nu}^2 
        + {1\over 2}{\rm Tr}\, (D_\mu\Phi)^2   -V(\Phi)  \, .
\end{equation}
Here
\begin{equation}
     D_\mu= \partial_\mu + i e A_\mu
\end{equation}
is the non-Abelian covariant derivative and
\begin{equation}
      F_{\mu\nu}=\partial_\mu A_\nu - \partial_\nu A_\mu 
             + ie [A_\mu, A_\nu]
\end{equation}
is the field strength with magnetic components $B_i=(1/2)\epsilon_{ijk}
F_{jk}$ and electric components $E_i=F_{oi}$. For definiteness, let us assume
that the potential is of the form  \begin{equation}
    V(\Phi) = - {\mu^2 \over 2} {\rm Tr}\,\Phi^2 
       +{\lambda\over 4} ({\rm Tr}\,\Phi^2)^2   \, .
\end{equation}

      If $\mu^2 <0$, the classical energy has a minimum at $\Phi=0$
that preserves the $SU(2)$ symmetry.  The spectrum of the quantum
theory includes three massless gauge bosons and three massive scalars
with equal masses.  If instead $\mu^2>0$, there is a family of
physically equivalent degenerate minima given by ${\rm Tr}\,\Phi_0^2 =
\mu^2/\lambda \equiv v^2$; the ``vacuum manifold'' of such minima can
be identified with the coset space $SU(2)/U(1)= S^2$.
In each of these vacuum states the
$SU(2)$ symmetry is spontaneously broken to the $U(1)$ subgroup that
leaves $\Phi_0$ invariant; this $U(1)$ subgroup may be identified with
electromagnetism.  After quantization of the theory, the small
fluctuations about the vacuum lead to a spectrum of elementary
particles that includes a massless photon, an electrically neutral
Higgs scalar with mass $\sqrt{2} \mu$, and a pair of vector bosons
with mass $ev$ and electric charges $\pm e$.

      In describing either the vacuum or configurations that are small
perturbations about the vacuum, it is most natural to take the
orientation of the Higgs field to be uniform in space; indeed, our
ansatz for the monopole solution corresponds to a vacuum with
$\Phi_0^a = v \delta^{a3}$.  However, the orientation of the Higgs
field is gauge-dependent quantity that need not be uniform.  In
particular, by applying a spatially varying $SU(2)$ gauge
transformation (with a singularity along the negative $z$-axis), we
can bring our monopole solution into the manifestly nonsingular
``radial gauge'' form
\begin{eqnarray}
     A_j^a &=& \epsilon_{jak} {\hat r}_k {1- u(r) \over er}  \cr
      \Phi^a &=& {\hat r}_a h(r)   \, .
\end{eqnarray}
At large-distances the resulting magnetic field  
\begin{equation}
   B_i^a ={1 \over e} {{\hat r}_i {\hat r}_a \over r^2} + O(1/r^3)  
\end{equation}
is parallel to $\Phi$ in internal space, showing that it lies in the unbroken
electromagnetic subgroup.

    In any finite energy solution, the Higgs field must approach one
of the minima of $V(\Phi)$ as $r \rightarrow \infty$ in any fixed
direction.  Hence, for any nonsingular solution the Higgs
configuration gives a map from the $S^2$ at spatial infinity into the
vacuum manifold $SU(2)/U(1)$.  Any such map corresponds to an element
of the homotopy group $\Pi_2(SU(2)/U(1)) = Z$ and can therefore be
assigned an integer ``topological charge'' $n$.  While the vacuum
solutions correspond to the identity element with $n=0$, the radial
gauge monopole solution gives a topologically nontrivial map with
$n=1$.  In fact, one can show that there is a one-to-one correspondence
between the magnetic charge and the topological charge, with a
nonsingular $SU(2)$ configuration of total magnetic charge $Q_M = 4\pi
n/e$ having topological charge $n$.  Although magnetic charges
that are half-integer multiples of $4\pi n /e$ are allowed by the
Dirac quantization condition, they cannot be obtained from nonsingular
field configurations.

\section{The BPS limit}
An especially interesting special case, which I will assume for the
remainder of these talks, is known as the
Bogomolny-Prasad-Sommerfield, or BPS, limit \cite{bps}.  It can be motivated by
considering the expression for the energy of a static configuration
with magnetic, but not electric, charge.  Assuming for the moment that
$A_0$ vanishes identically, we have
\begin{eqnarray}
    E &=& \int d^3x \left[ {1\over 2}{\rm Tr}\, B_i^2 
              +{1\over 2}{\rm Tr}\, (D_i\Phi)^2 +V(\Phi) \right]  \cr
       &=&  \int d^3x \left[ {1\over 2}{\rm Tr}\,(B_i \mp D_i\Phi)^2 +V(\Phi)
                \pm {\rm Tr}\, B_i D_i\Phi \right]     \, . 
\end{eqnarray}
With the aid of the Bianchi identity $D_iB_i =0$ the last term on the
right hand side may be rewritten as a surface integral over the sphere at
spatial infinity:
\begin{equation}
     \int d^3x {\rm Tr}\,  B_i D_i\Phi =  \int d^3x  \partial_i
    ({\rm Tr}\,B_i\Phi) = \int dS_i {\rm Tr}\, B_i\Phi   \equiv Q_M  v  \, .       
\end{equation}
(The normalization of $Q_M$ implied by the last equality agrees
with that of Eq.~(\ref{bfield}).)
Substituting this back into the previous equation yields the bound
\begin{eqnarray}
  E &=&  \pm Q_M v  + \int d^3x 
     \left[ {1\over 2}{\rm Tr}\,(B_i \mp D_i\Phi)^2 +V(\Phi) \right] \cr
    & \ge & |Q_M| v   \, .
\label{BPSbound}
\end{eqnarray}

The BPS limit is obtained by dropping the contribution of $V(\Phi)$ to
the energy.  This can be done most simply by letting $\mu^2
\rightarrow 0$, $\lambda \rightarrow 0$, with $v^2 = \mu^2/\lambda$
held fixed.  It can also be obtained by considering the extension of
this theory to a Yang-Mills theory with extended supersymmetry.  The
latter approach is particularly attractive from a physical point of
view, and can be formulated in such a way that the BPS limit is
preserved by higher order quantum corrections.

Now recall that any static configuration that is a local minimum of the
energy is a stable solution of the classical equations of motion.
Because the magnetic charge is quantized, any configuration that
saturates the lower bound in Eq.~(\ref{BPSbound}) will be such a solution.
With $V(\Phi)$ absent, the conditions for saturation of this bound are
the BPS equations
\begin{equation}
     B_i =D_i \Phi  \, .
\end{equation}
(I have assumed here, and henceforth, that $Q_M \ge 0$; the extension to the
case $Q_M <0$ is obvious.)  One can easily verify by direct substitution that
any solution of the first-order BPS equations is indeed a solution of the
second-order Euler-Lagrange equations. 

This result can easily be extended to the case of dyons, solutions carrying
not only a magnetic charge $Q_M$ but also a nonzero electric charge 
\begin{equation}
      Q_E  =  v^{-1}\int dS_i {\rm Tr}\, E_i\Phi \, .
\end{equation}
The bound on the energy is generalized to 
\begin{equation}
     E \ge v \sqrt{Q_M^2 + Q_E^2}
\end{equation}
with the minimum being achieved by configurations that satisfy
\begin{eqnarray}
     B_i &=& \cos \beta  D_i \Phi \cr
     E_i &=& \sin \beta  D_i \Phi \cr
     D_0 \Phi &=& 0
\end{eqnarray}
with $\beta = \tan^{-1}(Q_E/Q_M)$.

An attractive feature, which was in fact one of the original
motivations for the BPS approximations, is that it is possible to
obtain a simple analytic expression for the singly charged monopole
solution.  By a rescaling of fields and distances the gauge coupling $e$ can be
set equal to unity; for the remainder of these talks I will assume that this
has been done.   The solutions can then be written as \cite{bps},
\begin{eqnarray} 
    A_j^a &=& \epsilon_{jak} {\hat r}_k \left[ {v \over \sinh(vr)} 
      - {1\over r} \right] \cr 
    \Phi^a &=& {\hat r}_a \left[ v \coth (vr) - {1\over r} \right] \, .
\label{SUtwoBPS}
\end{eqnarray}
Note that the Higgs field does not approach its asymptotic value
exponentially fast, but instead has a $1/r$ tail.  This is because the
absence of a potential term makes the Higgs field massless.  Since a
massless scalar field carries a long-range force that is attractive
between like objects, this raises the possibility that the magnetic
repulsion between two BPS monopoles might be exactly cancelled by
their mutual scalar attraction, thus allowing for the existence of
static multimonopole solutions.  In fact, it turns out that such
solutions --- indeed continuous families of solutions --- exist for
all values of $Q_M$.

The actual construction of these multimonopole solutions is a
difficult, but fascinating, problem.
For the moment, I will simply concentrate on the problem of counting
the number of physically meaningful parameters, or ``collective
coordinates'', needed to specify these solutions.  Each of these
corresponds to a zero frequency eigenmode (a ``zero mode'') in the
spectrum of small fluctuations about a given solution.  However, there
are also an infinite number of zero modes, corresponding to local
gauge transformations of the solution, that do not correspond to any
physically meaningful parameter.  To eliminate these, a gauge
condition must be imposed on the fields.

I will start with the zero modes about the solution with
unit magnetic charge.  The elimination of the gauge modes is
particularly transparent if we work in the singular ``string gauge''
of Eq.~(\ref{stringgauge}) where $\Phi^1=\Phi^2=0$.  This leaves only a $U(1)$
gauge freedom that can be fixed by imposing, e.g, the electromagnetic
Coulomb gauge condition ${\bf \nabla \cdot {\cal A}}=0$.  Explicit solution
of the zero mode equations then shows that there are precisely four
normalizable zero modes about the solution.  Three of these correspond
to infinitesimal spatial translations of the monopole; the
corresponding parameters are most naturally chosen to be the spatial
coordinates of the center of the monopole.  The fourth zero mode
corresponds to a spatially constant phase rotation of the massive
vector field, $W_\mu ({\bf r}) \rightarrow e^{i\alpha} W_\mu ({\bf
r})$.  Since this mode is in fact a gauge mode that has no effect on
gauge-invariant quantities, one might think that it should be
discarded as unphysical.  The justification for not doing so comes
from considering the effect of allowing the collective coordinates to
be time-dependent.  In the case of the translation modes, this gives
a solution with nonzero linear momentum.  For the gauge mode, allowing
the phase $\alpha$ to vary linearly in time produces a dyon solution
that carries an electric charge proportional to $d\alpha/dt$.

Although explicit solution of the zero mode equations suffices for the
case of unit magnetic charge, where the monopole solution is known
explicitly, index theory methods are needed to count the zero modes
about solutions with higher charges \cite{SUtwoindex}.  Each zero mode
consists of perturbations $\delta A_j$ and $\delta \Phi$ that can be
viewed as three-component vectors transforming under the adjoint
representation of $SU(2)$.  Since these preserve the BPS equations,
they must satisfy
\begin{eqnarray}
     0 &=& \delta(B_j -D_j\Phi)  \cr
      &=& D_j\delta\Phi - \phi \delta A_j - \epsilon_{jkl} D_k \delta A_l \, .
\label{zeromode}
\end{eqnarray}
(Here $D_j= \partial_j + A_j$ and $A_j$ and $\Phi$ are  
$3\times 3$ anti-Hermitian matrices in the adjoint representation of $SU(2)$.)
These must be supplemented by a gauge condition that eliminates the
unwanted gauge modes.  A convenient choice is the background gauge
condition
\begin{equation} 
     0 = D_j\delta A_j +  \Phi \delta \Phi
\label{background}
\end{equation}
which is equivalent to requiring that the perturbation be
orthogonal, in the functional sense, to all normalizable gauge modes.
The number of collective coordinates is just equal to the number of
linearly independent normalizable solutions of Eqs.~(\ref{zeromode}) and
(\ref{background}).

If we define \cite{lowell}
\begin{equation}
    \psi = I \delta \Phi + i \sigma_j \delta A_j
\end{equation}
where $I$ is the unit $2\times 2$ matrix and the $\sigma_j$ are the Pauli
matrices,  Eqs.~(\ref{zeromode}) and (\ref{background}) can  be combined into
the single Dirac-type equation
\begin{equation}
    0 = (-i \sigma_j D_j + i \Phi) \psi \equiv {\cal D} \psi \, .
\label{Diraceq}
\end{equation}
We must remember, however, that two solutions $\psi$ and $i\psi$ that
are linearly dependent as solutions of Eq.~(\ref{Diraceq}) actually correspond
to linearly independent solutions of the original bosonic equations 
(\ref{zeromode}) and (\ref{background}).  The number of collective coordinates
is thus actually twice the number of linearly independent normalizable zero
eigenmodes of $\cal D$.  

Note that if $\psi({\bf r})$ is a solution of Eq.~(\ref{Diraceq}), then so is
\begin{equation}
     \psi'({\bf r})  = \psi({\bf r}) U
\label{rightmult}
\end{equation}
where $U$ is any $2\times 2$ unitary matrix.  This fact, which be of
importance later, implies that number of normalizable zero eigenmodes of the
bosonic equation must be a multiple of four.

The next step is to define
\begin{equation}
   {\cal I}(M^2) = {\rm Tr}\, {M^2 \over {\cal D}^\dagger {\cal D}+M^2} 
      - {\rm Tr}\, {M^2 \over {\cal D}{\cal D}^\dagger+M^2 }
\end{equation}
where $M$ is an arbitrary real number and 
\begin{equation}
   {\cal D}^\dagger  = -i \sigma_j D_j - i \Phi
\end{equation}
is the adjoint of $\cal D$.  The quantity
\begin{equation}
    {\cal I} = \lim_{M^2 \rightarrow 0} {\cal I}(M^2) 
\end{equation}
is then equal to the number of zero eigenvalues of ${\cal D}^\dagger
{\cal D}$ minus the number of zero eigenvalues of ${\cal D}{\cal
D}^\dagger$.  Using the fact that the unperturbed solution obeys the
BPS equations, one finds that
\begin{eqnarray}
   {\cal D}^\dagger {\cal D} &=& -D_j^2 + 2\sigma_j B_j +\Phi^2  \cr
    {\cal D}{\cal D}^\dagger  &=& -D_j^2  +\Phi^2  \, .
\end{eqnarray}
The second equation shows that ${\cal D}{\cal D}^\dagger$ is a
positive operator with no normalizable zero modes.  Since every
normalizable zero mode of $\cal D$ is also a normalizable zero mode of
${\cal D}^\dagger {\cal D}$, and conversely, $\cal I$ would clearly
give the desired counting of zero modes if it were not for the fact
that these operators have continuous spectra extending down to zero.

The contribution from these continuous spectra can be written as  
\begin{equation}
    {\cal I}_{\rm continuum} = \lim_{M^2\rightarrow 0} 
        \int {d^3k\over (2\pi)^3 }{m^2 \over k^2 +M^2} 
        \left[ \rho_{{\cal D}^\dagger {\cal D}}(k) 
       - \rho_{{\cal D}{\cal D}^\dagger }(k) \right]
\end{equation}
where $\rho_{\cal O}(k)$ is the density of continuum eigenstates of an
operator $\cal O$.  This contribution can be nonzero only if these
density of states factors are singular near $k=0$.  For the case at
hand, one can show that this is not the case.  The essential
idea is that such singularities are determined by the large $r$
behavior of the potential terms in the operators.  Since ${\cal
D}^\dagger {\cal D} - {\cal D}{\cal D}^\dagger = 2\sigma_j B_j$, the
potentially dangerous behavior is associated with the long-range
behavior of the magnetic field.  But, up to exponentially small
corrections, the long-range part of the $B_j$ lies in the unbroken
$U(1)$ subgroup and so does not act on the massless components of the
fields, which also lie in this $U(1)$.  Since only these latter fields
have spectra that extend down to zero, ${\cal I}_{\rm continuum}$
vanishes.

Having eliminated the continuum contribution, let us now turn to the
evaluation of $\cal I$.  For this purpose it is convenient to adopt a
pseudo-four-dimensional notation and define a four-vector $V_\mu$ with
components $V_j = A_j$ for $j =1,2,3$ and $V_4=\Phi$.  Because this is
actually a three-dimensional space, $\partial_4=0$ and so $D_4 =
\Phi$. Similarly, $G_{ij}= F_{ij}$ while $G_{i4}=-G_{4i}=D_i \Phi$.
Finally, the Dirac matrices
\begin{equation}
   \gamma_k = \left( \matrix{ 0 & -i\sigma_k \cr i\sigma_k & 0}
 \right)  \qquad
   \gamma_4 = \left( \matrix{ 0 & I \cr I & 0} \right)  \qquad
   \gamma_5 = \left( \matrix{ I & 0 \cr 0 & I} \right)
\end{equation}
all anticommute with each other and all have square equal to unity.

With these definitions, we can write
\begin{equation}
      \gamma_\mu D_\mu = \left( \matrix{ 0 & {\cal D} \cr -{\cal
 D}^\dagger  & 0}
          \right)
\end{equation}
and hence
\begin{equation}
   {\cal I}(M^2) = -{\rm Tr} \, \gamma_5 {M^2 \over -(\gamma\cdot D)^2 +M^2}
     = - \int d^3x \langle x|{\rm tr}\, \gamma_5 {M\over \gamma\cdot D + M}
          | x \rangle  \, .
\end{equation}
(Here Tr indicates a functional trace, while tr denotes a
trace over Dirac and $SU(2)$ matrix indices.)  The integrand in the last
expression can be written as the three-dimensional divergence of the current
\begin{equation}
    J_i = {1\over 2} \int d^3x \langle x|{\rm tr}\, \gamma_5 \gamma_i 
       {1\over \gamma\cdot D + M}
          | x \rangle  
     = - {1\over 2} \int d^3x \langle x|{\rm tr}\, \gamma_5 \gamma_i 
    (\gamma\cdot D) {1 \over -(\gamma\cdot D)^2 +M^2}| x \rangle
\end{equation}
and so 
\begin{equation}
   {\cal I}(M^2) = \int d^3x \partial_i J_i(x) = \int dS_i J_i (x)
\end{equation}
where the surface integral in the last term is over the sphere at spatial
infinity.  

We now write
\begin{equation}
   {1 \over -(\gamma\cdot D)^2 +M^2} = {1 \over -{\bf D}^2 +\Phi^2 +M^2 }
       + {1 \over -{\bf D}^2 +\Phi^2 +M^2 } 
     \left({i\over 2}\gamma_\mu \gamma_\nu G_{\mu\nu} \right) 
     {1 \over -{\bf D}^2 +\Phi^2 +M^2 } + \cdots
\end{equation}
where ${\bf D}^2 = D_jD_j$ and
the dots represent terms of order $G^2$ or higher that vanish at
least as fast as $1/r^4$ at spatial infinity.  When this expansion is
substituted into the expression for $J_i$, the contribution from the
first term vanishes after the trace over Dirac indices is performed.
The remaining terms give
\begin{eqnarray}
   {\hat x}_i J_i &=& -{i \over 2} \epsilon_{i\lambda\mu\nu} {\hat x}_i 
    {\rm tr}\, \langle x| D_\lambda {1 \over -{\bf D}^2 +\Phi^2 +M^2 }
     G_{\mu\nu} {1 \over -{\bf D}^2 +\Phi^2 +M^2 } |x\rangle 
      + O(x^{-4})  \cr
   &=& {i \over 2}  \langle x| \Phi {1 \over -{\bf \nabla}^2 +\Phi^2 +M^2 }
         {\hat x \cdot B} {1 \over -{\bf \nabla}^2 +\Phi^2 +M^2 } |x \rangle
      + O(|x|^{-3})
\end{eqnarray}
where the trace is now only over $SU(2)$ indices.  

The evaluation of this last expression is most transparent if we work in
the singular ``string gauge''.  If the magnetic charge is $Q_M=4\pi
n$, then asymptotically $\Phi \rightarrow v T^3$, and ${\hat x \cdot
B} \rightarrow (n/x^2) T^3$ and one finds that
\begin{equation}
  {\hat x}_i J_i = {n \over 2\pi x^2} {v\over \sqrt{v^2 +M^2}} 
       +  O(|x|^{-3}) \, .
\end{equation}
It follows that
\begin{equation}
   {\cal I}(M^2) = 2n {v\over \sqrt{v^2 +M^2}}
\label{almostI}
\end{equation}
and that the number of linearly independent normalizable zero modes of the
original bosonic problem is
\begin{equation}
    2 {\cal I} = 4n  \, .
\label{SUtwoI}
\end{equation}

A priori, one might have expected that classical
solutions with higher charges
could lead to new types of magnetically charged particles.
Eq.~(\ref{SUtwoI}), together with the fact that the BPS energy is
strictly proportional to the magnetic charge, suggests that this is
not the case.  Instead, all higher charged solutions should be viewed
as being multimonopole solutions composed of $n>1$ unit monopoles,
each with three translational and one $U(1)$ degree of freedom.\footnote{As was
the case with the unit monopole, there is only a single gauge mode that is not
eliminated by the gauge condition; this corresponds to a simultaneous $U(1)$
rotation of all the monopoles.  The modes corresponding to relative $U(1)$
rotations are not simply gauge transformations of the underlying solution.}
In the quantum theory, these solutions would thus correspond to
multiparticle states.

It is useful at this point to review the spectrum of particles in this
theory.  Quantization of the small fluctuations of the fundamental
fields yielded two particles, the photon and the Higgs scalar, that
have neither electric nor magnetic charge and that in the BPS limit
are both massless.  It also gave two massive vector particles, with
electric charges $\pm e$, no magnetic charge, and mass $ev$.  In
addition to these we have the monopole and antimonopole, with no
electric charge, magnetic charges $\pm 4\pi/e$, and mass $(4\pi/e)v$.
A curious feature of this spectrum is that the pattern of masses and
charges remains the same under the interchanges $e \leftrightarrow
4\pi/e$ and $Q_E \leftrightarrow Q_M$.  There is a mismatch in spin,
since the monopole and antimonopole are spinless, while the vector
bosons have spin one, but this can be remedied by enlarging the theory
so that it has $N=4$ extended supersymmetry \cite{osborn}; once this
is done, the elementary electrically charged particles and the
magnetically charged BPS soliton states form supermultiplets with
corresponding spins.  These facts suggest that this duality symmetry,
which exchanges solitons and elementary particles, and weak and strong
coupling, might in fact be an exact symmetry of the theory, as was
first conjectured by Montonen and Olive \cite{dual}.

\section{Monopoles in theories with larger gauge groups}

This analysis can be extended to the case of a Yang-Mills theory with an
arbitrary simple gauge group $G$ of rank $r$ and dimension $d$ and a 
Higgs field $\Phi$ transforming under the adjoint representation.  To begin,
recall that the generators of the Lie algebra of $G$ can be chosen to be $r$
commuting generators $H_a$ that span the Cartan subalgebra, together
with a number of generators $E_\balpha$ associated with the $d-r$ root
vectors $\balpha$ that are defined by the commutation relations
\begin{equation}
    [E_\balpha , H_j ] =  \alpha_j E_\balpha  \, .
\end{equation}

The asymptotic value of the Higgs field in some fixed reference 
direction can always be chosen to lie in the Cartan subalgebra.  It  
thus defines an $r$-component vector $\bf h$ through the relation  
\begin{equation}
      \Phi_0 = {\bf h} \cdot {\bf H}  \, .
\end{equation}
The unbroken gauge symmetry is the subgroup $G$ that leaves
$\Phi_0$ invariant.  The maximal symmetry breaking occurs if $\bf h$
has nonzero inner products with all the root vectors, in which case
the unbroken subgroup is the $U(1)^r$ generated by the Cartan
subalgebra.  If instead some of the roots are orthogonal to $\bf h$,
then these form the root lattice for a non-Abelian group $K$ of rank
$k<r$ and the unbroken symmetry is $U(1)^{r-k} \times K$.

At large distances, $F_{\mu\nu}$ must commute with the Higgs field.
Hence, along the same direction used to define $\bf h$, the asymptotic
magnetic field may be chosen to also lie in the Cartan subalgebra and
to be of the form
\begin{equation}
      B_i = {\bf g} \cdot {\bf H} {{\hat r}_i\over r^2} + O(r^{-3})  \, .
\end{equation}
The generalized quantization condition on the magnetic charge then becomes
\cite{topology}
\begin{equation}
      e^{i{\bf g}\cdot {\bf H}} = I  \, .
\label{generalquant}
\end{equation}
 
I will begin by considering the case of maximal symmetry breaking.
Because $\Pi_2(G/U(1)^r) = \Pi_1(U(1)^r) = Z^r$, there are $r$
topologically conserved charges.  These can be identified in a
particularly natural fashion by recalling that a basis for the root
lattice can be chosen to be a set of $r$ simple roots $\bbeta_a$
with the property that all other roots are linear combinations of  
simple roots with coefficients that are either all positive or all
negative.  There are many possible choices for this basis.  However, a
unique set of simple roots can be specified by requiring that
\begin{equation}
     {\bf h} \cdot \bbeta_a \ge 0
\label{simplechoice}
\end{equation}
for all $a$.  If all of the fields are in the adjoint representation, 
the quantization condition
(\ref{generalquant}) then reduces to the requirement that
\begin{equation}
{\bf g} = {4\pi} \sum_a n_a {\bbeta_a}^*
\end{equation}
where ${\bbeta_a}^* = \bbeta_a/ \bbeta_a^2$ and the integers $n_a$ are 
are the topological charges.  

The BPS mass formula is easily extended to this case.  One finds that 
\begin{equation}
      M = {\bf g}\cdot {\bf h} 
     = \sum_a n_a \left( {4\pi\over e} {\bf h} \cdot \bbeta_a  \right)
     \equiv \sum_a n_a m_a  \, .
\end{equation} 

The methods used to count the zero modes about $SU(2)$ solutions can
also be applied here \cite{erick}.  As before, there is no continuum
contribution ${\cal I}_{\rm continuum}$ because the long-range part of
the magnetic field lies in the Cartan subalgebra and so does not act
on the massless fields, which also lie in the Cartan subalgebra.  The
calculation of $\cal I$ proceeds very much as before until one gets to
Eq.~(\ref{almostI}), which is replaced by
\begin{equation}
     {\cal I}(M^2) = {1\over 4 \pi} \sum_{\balpha} 
      { (\balpha\cdot{\bf h}) (\balpha\cdot{\bf g}) 
       \over \left[ (\balpha\cdot{\bf h})^2 + M^2\right]^{1/2} }
       = {1\over 2 \pi} {\sum_{\balpha}}' 
      { (\balpha\cdot{\bf h}) (\balpha\cdot{\bf g}) 
       \over \left[ (\balpha\cdot{\bf h})^2 + M^2\right]^{1/2} } \, .
\end{equation}
Here the first sum is over all roots $\balpha$, while the prime on the
second sum indicates that it is to be taken only over the positive
roots (those that are positive linear combinations of simple
roots).  Taking the limit $M^2 \rightarrow 0$ gives
\begin{equation}
    {\cal I} = {1\over 2 \pi} {\sum_{\balpha}}' \balpha\cdot{\bf g}
       = 2 \sum_a n_a \left({\sum_\balpha}' \balpha\cdot
       {\bbeta_a}^*\right) \, .
\end{equation}
In the sum inside the parentheses, the contributions from the roots
other than $\bbeta_a$ cancel, so that the sum is just $\bbeta_a \cdot
{\bbeta_a}^*=1$.  Hence, the number of normalizable zero modes is
\begin{equation}
     2 {\cal I} = 4 \sum_a n_a \, .
\label{twoI}
\end{equation}

It was argued above that the $SU(2)$ solutions with higher magnetic
charge should be understood as being composed of a
number of unit monopoles.  The mass formula and the zero mode counting
suggest that the higher charged solutions in the present case should
also be understood as multimonopole solutions.  Now, however, there
are $r$ different species of fundamental monopoles, with the $a$th
fundamental monopole having mass $m_a$, topological charges $ n_b =
\delta_{ab}$ and four degrees of freedom.  Classical solutions
corresponding to these fundamental monopoles can be constructed by
appropriate embeddings of the $SU(2)$ solution.  Any root $\balpha$
defines an $SU(2)$ subgroup of of $G$ with generators
\begin{eqnarray}
t^1({\mbox{\boldmath $\alpha$}}) &=& \frac{1 }{ \sqrt{2
{\mbox{\boldmath $\alpha$}}^2}} (E_{\mbox{\boldmath $\alpha$}} +
E_{-{\mbox{\boldmath $\alpha$}}})                  \nonumber \\
t^2({\mbox{\boldmath $\alpha$}}) &=& -\frac{i }{ \sqrt{2{
\mbox{\boldmath $\alpha$}}^2}} (E_{\mbox{\boldmath $\alpha$}} -
E_{-{\mbox{\boldmath $\alpha$}}})                 \nonumber \\
t^3({\mbox{\boldmath $\alpha$}}) &=&  {\mbox{\boldmath $\alpha$}}^* 
\cdot  {\bf H} \, .
\label{tripletdef}
\end{eqnarray}
If we denote by 
$A^s_i({\bf r}; v)$ and $\Phi^s({\bf r}; v)$ the
unit $SU(2)$ monopole with Higgs expectation value $v$, then the
embedded solution
\begin{eqnarray}
   A_i({\bf r})  &=&  \sum_{s=1}^3 A_i^s({\bf r}; {\bf h}\cdot 
{\mbox{\boldmath $\beta$}}_a) t^s({\mbox{\boldmath $\beta$}}_a) \nonumber \\
   \Phi({\bf r})  &=&  \sum_{s=1}^3 \Phi^s({\bf r}; {\bf h}\cdot 
{\mbox{\boldmath $\beta$}}_a) t^s({\mbox{\boldmath $\beta$}}_a)   
   + ({\bf h} - {\bf h}\cdot {\mbox{\boldmath $\beta$}}_a^* \,
{\mbox{\boldmath $\beta$}})\cdot {\bf H} 
\label{embedding}
\end{eqnarray}
gives the fundamental monopole corresponding to the root $\bbeta_a$.  It
has the expected mass and topological charges and four zero modes,
three corresponding to translational degrees of freedom and the fourth
to a phase angle in the $U(1)$ generated by $\bbeta_a \cdot {\bf H}$.

As an example, consider the case of $SU(3)$ broken to $U(1)\times
U(1)$ by an adjoint representation Higgs field that can be represented
by a traceless Hermitian $3\times 3$ matrix.  Let $\Phi_0$ be
diagonal, with its eigenvalues decreasing along the diagonal.  With
this convention, the $SU(2)$ subgroup defined by $\bbeta_1$ lies in the
upper left $2\times 2$ block.  Embedding the $SU(2)$ monopole in this
block gives a solution with a mass $m_1$, topological charges
$n_a=(1,0)$, and four zero modes. After quantization, there is a
family of monopole and dyon one-particle states corresponding to this
solution.  Similarly, $\bbeta_2$ defines an $SU(2)$ subgroup lying in
the lower right $2\times 2$ block.  Using this subgroup for the
embedding gives a solution with mass $m_2$, topological charges
$(0,1)$, and again four zero modes.  This, too, corresponds to a
particle in the spectrum of the quantum theory.

There is a third $SU(2)$ subgroup, lying in the four corner matrix
elements, defined by the composite root $\bbeta_1 + \bbeta_2$.  Using
this subgroup to embed the $SU(2)$ monopole also gives a spherically
symmetric BPS solution, with mass $m_1 +m_2$ and topological charges
$(1,1)$.  However, Eq.~(\ref{twoI}) (as well as explicit solution of
the zero mode equations) shows that there are not four, but eight zero
modes.  Hence, this embedding solution is just one out of a
continuous family of two-monopole solutions; in contrast to the two
fundamental solutions, it can be continuously deformed into a solution
containing two widely separated fundamental monopoles.  It does not
lead to a new particle in the spectrum of the quantum theory, but
instead corresponds to a two-particle state.

Let us now consider this result in the light of the Montonen-Olive
duality conjecture.  Although this conjecture was first motivated by the
spectrum of the $SU(2)$ theory, it is natural to test it with larger gauge
groups.  The elementary particle sector of the theory contains a
number of massless particles, carrying no $U(1)$ charges, that are
presumably self-dual.  There are also six massive vector bosons, one
for each root of the root diagram, that carry electric-type charges in
one or both of the unbroken $U(1)$'s.  The duals of the $\pm \bbeta_1$
and $\pm \bbeta_2$ vector bosons are clearly the one-particle states
corresponding to the $\bbeta_1$- and $\bbeta_2$-embeddings of the
$SU(2)$ monopole and antimonopole solutions.  One might have thought
that the duals of the vector bosons corresponding to $\pm (\bbeta_1
+\bbeta_2)$ would be obtained from the $(\bbeta_1
+\bbeta_2)$-embedding solutions, but we have just seen that these do
not correspond to single-particle states.  Some other state must be
found if the duality is to hold.  The most likely candidate 
would be some kind of threshold bound state \cite{sen}.  To explore this
possibility, we need to understand the interactions of low-energy BPS
monopoles.  This can be done by making use of the moduli space
approximation, to which I now turn.

\section{The moduli space approximation}
The essential idea of the moduli space approximation \cite{manton} is
that, since the static multimonopole solutions are all BPS, the
time-dependent solutions containing monopoles with sufficiently small
velocities should in some sense also be approximately
BPS.\footnote{Here velocities should be understood to include not only
spatial velocities but also the time derivatives of the $U(1)$ phases.
Thus, we are considering slowly moving dyons with small (and possibly
zero) electric charges.}

To make this more precise, let $\{A_i^{\rm BPS}({\bf r}, z),
\Phi^{\rm BPS}({\bf r}, z)\}$ be a family of static, gauge-inequivalent
BPS solutions parameterized by a set of collective coordinates $z_j$.
The moduli space approximation is obtained by assuming that the fields
at any fixed time are gauge-equivalent to some configuration in this
family, so that they can be written as
\begin{eqnarray}
     A_0({\bf r},t) &=& 0 \cr 
     A_i({\bf r},t) &=& U^{-1}({\bf r},t)  A_i^{\rm BPS}({\bf r}, z(t)) U(({\bf
       r},t) -{i} U^{-1}({\bf r},t) \partial_i U(({\bf r},t) \cr
       \Phi({\bf r},t) &=& U^{-1}({\bf r},t) \Phi^{\rm BPS}({\bf r}, z(t))
      U(({\bf r},t) \, .
\label{modspaceansatz}
\end{eqnarray}
Their time derivatives are then of the form
\begin{eqnarray}
    \dot A_i &=& \dot z_j \left[{\partial A_i \over \partial z_j} + D_i
           \epsilon_j \right]  \equiv \dot z_j  \delta_j A_i  \cr
    \dot \Phi &=& \dot z_j \left[{\partial \Phi \over \partial z_j} + [\Phi,
           \epsilon_j] \right]  \equiv \dot z_j  \delta_j \Phi  
\end{eqnarray}
where the gauge function $\epsilon_j({\bf r},t)$ arises from the time
derivative of $U({\bf r},t)$.  These are constrained by Gauss's law,
which takes the form
\begin{eqnarray}
     0 &=& -D_\mu F^{\mu 0} + [\Phi, \partial_0\Phi] 
            = D_i\dot A_i + [\Phi, \dot \Phi]   \cr
       &=&  \dot z_j \left(D_i \delta_j A_i 
         + [\Phi, \delta_j \Phi] \right)  \, .
\end{eqnarray} 
Because they arise from variation of a collective coordinate, the
quantities $\delta_j A_i$ and $\delta_j \Phi$ form a zero mode about
the underlying solution BPS solution.  The Gauss's law constraint
shows that they obey the background gauge condition Eq.~(\ref{background}).

With $A_0$ identically zero, the Lagrangian of the theory can be written as
\begin{equation}
    L = {1\over 2} \int d^3r {\rm Tr}\left[ {\dot A_i}^2 + {\dot
          \Phi}^2 + B_i^2 + 
          D_i\Phi^2 \right] \, .
\end{equation}
Since for fields obeying the ansatz (\ref{modspaceansatz}) the
configuration at any fixed time is BPS, the contribution of the last
two terms to the integral is just the BPS energy determined by the
topological charge.  This is a time-independent constant that has no
effect on the dynamics and so can be dropped.  The remaining terms
then give an effective Lagrangian
\begin{equation}
    L_{\rm MS} = {1\over 2} g_{ij}(z) \dot z_i \dot z_j
\label{LMS}
\end{equation}
where 
\begin{equation}
    g_{ij}(z) = \int d^3r \left[ \delta_i A_k \delta_j A_k + 
        \delta_i \Phi  \delta_j \Phi  \right]  \, .
\label{metricdef}
\end{equation}
Thus, the full field theory dynamics for low energy monopoles has been
reduced to a problem involving a finite number of degrees of freedom.
If one views $g_{ij}(z)$ as a metric for the moduli space spanned by
the collective coordinates, the dynamics described by $L_{\rm MS}$ is
simply geodesic motion on the moduli space.

\section{Determining the moduli space metric}

Actually determining the moduli space metric is a nontrivial matter.
To apply Eq.~(\ref{metricdef}) directly one needs to know the zero modes,
whereas we do not in general even know the underlying solution.
However, some more indirect approaches can sometimes be brought to
bear on the problem.  

First, Gibbons and Manton \cite{gary} showed how one could obtain the metric
for the region of moduli corresponding to widely separated monopoles.  They
pointed out that, since the moduli space metric determines the low energy
dynamics, the metric can be inferred if this dynamics is known.  
The only long-range interactions between widely separated monopoles are
those mediated by massless fields.  These are the electromagnetic
interactions and an interaction due to the massless Higgs field.  The
Lagrangian describing the interactions between moving point electric
and magnetic forces is well known, while that for the scalar force is
easily worked out.  To obtain the metric, these must be expanded up to terms
quadratic in the velocities and the electric charges.  A Legendre
transformation must then be used to replace the electric charges by the time
derivatives of the corresponding phase angles.  Apart from a constant term, the
result is a Lagrangian, of the form of
Eq.~(\ref{LMS}), from which the metric can be read off directly.  For
the case of many $SU(2)$ monopoles, each with mass $m$ and magnetic charge
$g$ and with positions ${\bf x}_i$ and phase angles $\xi_i$,
this gives
\begin{equation}
   ds^2=\frac{1}{2}M_{ij}d{\bf x}_i\cdot d{\bf x}_j+\frac{g^4}{2(4\pi)^2} 
(M^{-1})_{ij}(d\xi_i+{\bf W}_{ik}\cdot d{\bf x}_k)(d\xi_j+{\bf W}_{jl}
\cdot d{\bf x}_l) 
\label{metric}
\end{equation}
where 
\begin{eqnarray}
M_{ii} &=& m  - \sum_{k\ne i} \frac{g^2}{ 4\pi r_{ik}},\nonumber \\
M_{ij} &=&\frac{g^2}{ 4\pi r_{ij}}\qquad
\hbox{\hskip 1cm if $i\neq j$},
\label{Mdef}
\end{eqnarray}  
and
\begin{eqnarray}
{\bf W}_{ii}&=&-\sum_{k\neq i}{\bf w}_{ik},\nonumber\\
{\bf W}_{ij}&=&{\bf w}_{ij}\qquad
\hbox{\hskip 1cm if $i\neq j$},
\label{Wdef}
\end{eqnarray}
with $r_{ij}$ the distance between the $i$th and $j$th monopoles and
${\bf w}_{ij}$ the value at ${\bf x}_i$ of the Dirac vector potential
due to the $j$th monopole, defined so that
\begin{equation}
    {\bf \nabla}_i \times {\bf w}_{ij}({\bf x}_i -{\bf x}_j)  = -
     \frac{ {\bf x}_i -{\bf x}_j }{ r_{ij}^3 }.
\end{equation}
           
     The extension of this result to the case of maximal symmetry
breaking of an arbitrary simple group $G$ is quite simple.  For a
collection of fundamental monopoles, with the $i$th monopole
corresponding to the simple root ${\mbox{\boldmath $\beta$}}_i$, we need
only replace Eqs.~(\ref{Mdef}) and (\ref{Wdef}) by
\begin{eqnarray}
M_{ii} &=& m_i  - \sum_{k\ne i} \frac{g^2 {\mbox{\boldmath $\beta$}}_i^* 
\cdot {\mbox{\boldmath $\beta$}}_k^*}{ 4\pi r_{ik}},\nonumber \\
M_{ij} &=&\frac{g^2 {\mbox{\boldmath $\beta$}}_i^* \cdot 
{\mbox{\boldmath $\beta$}}_j^*}{ 4\pi r_{ij}}\qquad
\hbox{\hskip 1cm if $i\neq j$},
\label{newM}
\end{eqnarray}
and 
\begin{eqnarray}
{\bf W}_{ii}&=&-\sum_{k\neq i}{\mbox{\boldmath $\beta$}}_i^*\cdot
{\mbox{\boldmath $\beta$}}_k^*{\bf w}_{ik},\nonumber\\
{\bf W}_{ij}&=&{\mbox{\boldmath $\beta$}}_i^*\cdot
{\mbox{\boldmath $\beta$}}_j^*{\bf w}_{ij}\qquad
\hbox{\hskip 1cm if $i\neq j$},
\end{eqnarray}
with $m_i=g\,\mbox{\boldmath $\beta$}_i^*\cdot {\bf h}$.

    Although the derivation of these expressions was only valid in the
region of moduli space corresponding to widely separated monopoles,
one might wonder whether the asymptotic metric could be exact.
For the case of two $SU(2)$ monopoles, several considerations show
that it cannot be.  The matrix $M$ of Eq.~(\ref{Mdef}) reduces to
\begin{equation}
    M= \left(\matrix{ m - {g^2\over 4\pi r}  & {g^2\over 4\pi r} \cr
         {g^2\over 4\pi r} & m - {g^2\over 4\pi r} }\right)   \, .    
\end{equation} 
The determinant of this matrix vanishes at $r=g^2/2\pi m$, implying a
singularity in the metric, despite the fact that there is no reason to
expect any type of singular behavior near this value of the
intermonopole distance.  Furthermore, we know that there is a
short-range force, carried by the massive vector bosons, that was
ignored in the derivation of the metric.  If one works in a singular
gauge in which the Higgs field orientation is uniform in space, this
interaction is proportional to the gauge-invariant quantity ${\rm
Re}[{\bf W}_{(1)}^* {\bf W}_{(2)}]$ where ${\bf W}_{(1)}$ and ${\bf
W}_{(2)}$ are the massive vector fields of the two monopoles.  Because
these fall exponentially with distance from the center of the
monopole, their overlap, and hence the interaction, falls
exponentially with the intermonopole separation.

Neither of these objections apply when the two monopoles are 
fundamental monopoles associated with 
different simple roots of a large gauge group.  The simple roots
have the property that their mutual inner products are always negative.  The
resulting sign changes in $M$ eliminate the zero of the determinant and make
the asymptotic metric everywhere nonsingular.  In addition, the
quantity characterizing the interactions carried by the massive vector
fields is now ${\rm Re}[{\rm Tr}{\bf W}_{(1)}^\dagger {\bf
W}_{(2)}]$, which vanishes when the two monopoles arise from different
simple roots.

This, of course, is not sufficient to show that the asymptotic metric
is exact.  To do this, we first note that the coordinates for the
moduli space can always be chosen so that three specify the position
of the center-of-mass of the monopoles and a fourth is an overall
$U(1)$ phase.  The moduli space metric can then be written in the
factorized form
\begin{equation}
    {\cal M} = R^3 \times \left({ R^1 \times {\cal M}_{\rm rel} \over D}
           \right ) 
\end{equation}
where the factors of $R^3$ and $R^1$ are associated with the center-of-mass
coordinates and the overall $U(1)$ phase, while ${\cal M}_{\rm rel}$ is the
metric on the subspace spanned by the relative positions and phases.  The
factoring by the discrete group $D$ arises from difficulties in globally
factoring out an overall $U(1)$ phase.   

${\cal M}_{\rm rel}$ has several important properties.  First, it must have a
rotational isometry reflecting the fact that the interactions among an
assembly of monopoles are unaffected by an overall spatial rotation of the
entire assembly.   Second, the $SU(2)$ relations among the zero modes
shown in Eq.~(\ref{rightmult}) imply that the moduli space metric must be
hyper-Kahler\footnote{A metric is hyper-Kahler if it possess three
covariantly constant complex structures that also form a quaternionic
structure and if it is pointwise Hermitian with respect to each.}
\cite{atiyah}.  Finally, the relative moduli space for a collection of $n$
monopoles is $4(n-1)$-dimensional.   Hence, we are seeking a four-dimensional
hyper-Kahler manifold with a rotational isometry.  There are four 
such \cite{atiyah}:

\noindent 1)  Flat four-dimensional Euclidean space  

\noindent 2)  The Eguchi-Hanson manifold \cite{eguchi}

\noindent 3)  The Atiyah-Hitchin manifold \cite{atiyah}

\noindent 4)  Taub-NUT space  

The first of these would imply that there were no interactions at all
between the monopoles, and so is clearly ruled out if $\bbeta_1 \cdot
\bbeta_2 \ne 0$.  The Eguchi-Hanson metric has the wrong
asymptotic behavior for large intermonopole separation, and so can be ruled out.
At large $r$ (but not at small $r$) the Atiyah-Hitchin metric
approaches the two-monopole asymptotic metric with $M$ given by
Eq.~(\ref{Mdef}).  It thus describes the moduli space for two $SU(2)$
monopoles (or for two identical monopoles in a larger group), but
not that for two distinct monopoles.  The only remaining possibility is
the Taub-NUT metric.  This not only agrees at large $r$ with the
asymptotic metric, but is in fact equal to it everywhere.  Thus,
for the case of two distinct fundamental monopoles the asymptotic metric
is in fact exact \cite{klee,gaunt,connell}.

If a collection of more than two monopoles includes two corresponding
to the same simple root, then the asymptotic metric develops a
singularity when these approach each other.  However, this metric is
everywhere nonsingular if the monopoles are all distinct.  It is
therefore natural to conjecture that for this case also the asymptotic
metric is exact \cite{klee2}.  Proofs of this conjecture have been given by
Chalmers \cite{chalmers} and by Murray \cite{murray}.

Let us now briefly return to the issue of duality in the theory with $SU(3)$
broken to $U(1)\times U(1)$.   As noted above, duality is expected to hold
only if the theory has an extended supersymmetry, which means that the
low-energy fermion dynamics must be included.  It turns out that these
fermions will give rise to a supermultiplet of threshold bound states if and
only if there is a normalizable harmonic form on the relative moduli space
\cite{witten}.  Having determined the metric for this moduli space metric, one
can easily verify that such a harmonic form exists, and hence that the test of
the duality conjecture is met \cite{klee,gaunt}.

\section{Nonmaximal symmetry breaking}
Let us now turn to the case of non-maximal symmetry breaking, where
the gauge symmetry $G$ is spontaneously broken to $K \times
U(1)^{r-k}$.  As in the case of maximal symmetry breaking, we can
require that inner products of the simple roots with $\bf h$ be all
non-negative.  It is useful to distinguish between those for which
this inner product is greater than zero and those for which it
vanishes.  I will continue to denote the former by $\bbeta_a$, and
will label the latter, which form a set of simple roots for $K$, by
$\bgamma_i$.  In contrast with the previous case, the condition on the
inner products with $\bf h$ does not uniquely determine the set of
simple roots.  Instead, there can be many acceptable sets, all related
by Weyl reflections of the root diagram that result from global gauge
transformations by elements of $K$.\footnote{Consider, for example,
the case of $SU(3)$ broken to $SU(2)\times U(1)$.  If one set of
simple roots is denoted by $\beta$ and $\gamma$, with the latter
being a root of the unbroken $SU(2)$, then another acceptable set is
given by $\beta +\gamma$ and $-\gamma$.}

The quantization condition on the magnetic charge now takes the form
\begin{equation}
{\bf g} = {4\pi} \left[ \sum_a n_a {\bbeta_a}^* + 
          \sum_j q_j{\bgamma_j}^* \right]  \, .
\end{equation}
As in the case of maximal symmetry breaking, the integers $n_a$ are the
topological charges, one for each $U(1)$ factor of the unbroken group.  They
are gauge-independent, and thus independent of the choice of the set of simple
roots.   The $q_j$ must also be integers, but they are neither
topologically conserved nor gauge-invariant.   We will see that there is an
important distinction to be made between the case where
\begin{equation}
      {\bf g} \cdot \bgamma_j =0 \, , \qquad {\rm all \,\, }j,
\label{purelyabelian}
\end{equation}
and that where some of the ${\bf g} \cdot \bgamma_j$ are nonzero.  (Note that
these do not in general correspond to vanishing or nonvanishing $q_j$.) In the
former case, the long-range magnetic fields are purely Abelian with only
$U(1)$ components, whereas in the latter the configuration has a non-Abelian
magnetic charge.  We will see that there are a number of pathologies associated
with the latter case.

The BPS mass formula takes the same form as before,
\begin{equation}
      M= \sum_a n_a m_a
\label{nonAmass}
\end{equation}
but with the sum running only over the indices corresponding to simple
roots that are not orthogonal to $\bf h$.

The zero mode counting proceeds as before, but with some complications
\cite{erick2}.  First, the continuum contribution cannot be
immediately discarded, since the massless fields cannot all be brought
into the Cartan subalgebra.  Because of this, there can be a
singularity in the density of states factor that is strong enough to
give a nonzero ${\cal I}_{\rm continuum}$ if ${\cal D}^\dagger {\cal
D} - {\cal D}{\cal D}^\dagger$ contains order $1/r^2$ terms that act
on fields lying in the unbroken non-Abelian subgroup $K$.  This will
be the case whenever there is a net non-Abelian magnetic charge.
Explicit solution of the zero mode equations in some simple cases
shows that the number of zero modes is not equal to the expression for
$2 {\cal I}$ given below, implying that there is indeed a nonvanishing
continuum contribution.  This difficulty does not arise when
Eq.~(\ref{purelyabelian}) is satisfied.

Second, the expression for $\cal I$ is more complicated.  The same procedures
as used before again lead to
\begin{equation}
     {2\cal I}= \lim_{M^2 \rightarrow 0}
           {1\over \pi} {\sum_{\balpha}}' 
      { (\balpha\cdot{\bf h}) (\balpha\cdot{\bf g}) 
       \over \left[ (\balpha\cdot{\bf h})^2 + M^2\right]^{1/2} }  
\label{nonAalmostI}
\end{equation}
with the prime indicating that the sum is only over positive roots.  Now,
however, the contribution from the  roots orthogonal to $\bf h$ (i.e.,
those of the subgroup $K$) vanishes even for finite $M^2$ and so gives no
contribution to the limit.   As a result, the expression for ${2\cal I}$ is in
general much less simple than before.  But, again, matters simplify if the
asymptotic magnetic field is purely Abelian.  Because the roots of $K$ are
now all orthogonal to $\bf g$, they would not have contributed in any case,
and the methods used for the case with maximal symmetry breaking yield
\begin{equation}
     {2\cal I}=  4 \left[ \sum_a n_a  + \sum_j q_j \right] \, .
\label{nonAtwoI}
\end{equation}
As was noted earlier, the $q_j$ are not gauge-invariant.  However, when $\bf g$
is orthogonal to all of the $\bgamma_k$, the sum appearing on the
right-hand side of Eq.~(\ref{nonAtwoI}), and hence $2 \cal I$, is
gauge-invariant.

The difficulties with applying index theory when there is a
non-Abelian magnetic charge are related to other known difficulties
with such solutions.  Since the unbroken gauge group acts nontrivially
on these, one would expect to find gauge zero modes, analogous to the
$U(1)$ modes of the maximally symmetric case, whose excitation would
lead to ``chromodyons'', objects with non-Abelian electric-type
charge.  Instead, one finds that these modes are non-normalizable and
that the expected chromodyon states are absent \cite{abouel}.  This
can be traced to the fact that the existence of the non-Abelian
magnetic charge creates an obstruction to the smooth definition of a
set of generators for $K$ over the sphere at spatial infinity; i.e.,
one cannot define ``global color'' \cite{manohar}.

It is instructive to return to the $SU(3)$ example considered in
Sec.~4, but with the last two eigenvalues of $\Phi_0$ taken to be
equal so that the unbroken group is $SU(2)\times U(1)$.  While $n_1$
remains a topological charge, $n_2$ must be replaced by the
nontopological integer $q_1$. The first fundamental monopole solution
of the maximally broken case, obtained by embedding in the upper left
$2\times 2$ block, is still present with a nonzero mass.  As before,
it has three translational zero modes and a $U(1)$ phase mode.  There
are no other normalizable zero modes, even though the solution is not
invariant under the unbroken $SU(2)$, and even though
Eq.~(\ref{nonAalmostI}) gives ${2\cal I}= 6$.  Embedding in the lower
right $2\times 2$ block, which previously gave a second fundamental
monopole, is no longer possible.  Indeed, if one starts with the
maximally broken case, and follows the behavior of the second
fundamental monopole as the last two eigenvalues of $\Phi_0$ approach
one another, one finds that its mass tends to zero, its core radius
tends to infinity, and the fields at any fixed point approach their
vacuum value.  Finally, the embedding in the corner matrix elements,
which previously gave a solution with eight zero modes that was
naturally understood to be a two-monopole solution, now gives a
solution that is gauge-equivalent to the first fundamental monopole
and hence has only four zero modes.  In all three of these cases the
magnetic charge has a non-Abelian component.

Eqs.~(\ref{nonAmass}) and (\ref{nonAtwoI}) are consistent with the
idea that even for non-maximal symmetry breaking one should interpret
all solutions --- or at least those with purely Abelian magnetic
charges --- in terms of a number of component fundamental monopoles.
However, there are clearly two quite different kinds of fundamental
monopoles. The massive monopoles corresponding to the $\bbeta_a$ carry
$U(1)$ magnetic charges and appear to have four associated degrees of
freedom. They can be realized as classical solitons, even though the
latter may not be unique, as the $SU(3)$ example shows.  The remaining
fundamental monopoles, corresponding to the $\bgamma_j$, would have to
be massless.  Indeed, the duality conjecture would lead us to expect
to find massless magnetically charged states that would be the duals
of the massless gauge bosons of the unbroken non-Abelian subgroup.
The difficulty is that, precisely because they are massless, these
monopoles cannot be associated with any localized classical solutions.
To learn more about them, we must examine multimonopole solutions
containing both massive and massless components.

The pathologies associated with non-Abelian magnetic charges suggest that this 
is best done by concentrating on configurations that obey
Eq.~(\ref{purelyabelian}).   This should not impose any real physical
restriction, since the additional monopoles needed to cancel any non-Abelian
charge can be placed at an arbitrarily large distance.  It also turns out to be
useful to treat non-maximal symmetry breaking as a limit of maximal symmetry
breaking in which one or more of the ${\bf h}\cdot \bbeta_a^*$ tend to zero.  As
we will see, it appears that the moduli space for the maximally broken case
behaves smoothly in this limit, with the limit of its metric being the metric
for the non-maximally case.  Although some of the fundamental monopoles become
massless in this limit and no longer have corresponding soliton solutions, their
degrees of freedom of these massless monopoles are still evident in the
low-energy moduli space Lagrangian.

\section{An $SO(5)$ example}

A particularly simple example \cite{nonabelian} for illustrating this
arises with the gauge group $SO(5)$, whose root diagram is shown in
Fig.~1 with the simple roots labeled $\bbeta$ and $\bgamma$.
Consider the solutions whose magnetic charge is such that
\begin{equation}
    {\bf g} = {4\pi } \left( \bbeta^* + \bgamma^* \right) \, .
\end{equation}
With $\bf h$ as in Fig.~1a, the symmetry breaking is maximal and there
is an eight-parameter family of solutions composed of two monopoles, of masses
$m_\bbeta$ and $m_\bgamma$ respectively.  Because the
two monopoles correspond to different simple roots, the moduli space
metric is known from the results 

\vskip 5mm
\begin{center}
\leavevmode
\epsfysize =2in\epsfbox{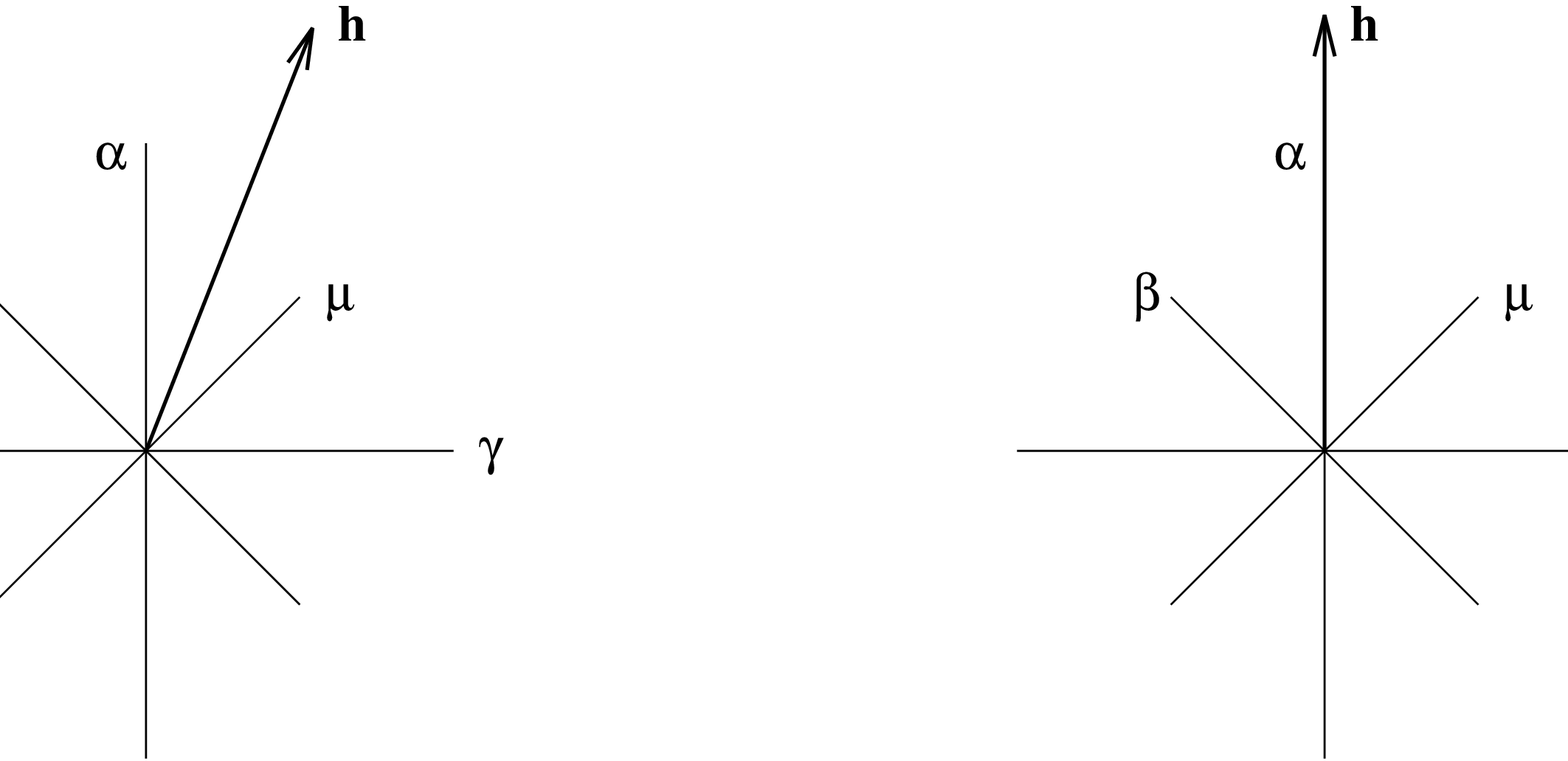}
\end{center}
\vskip 8mm
\begin{quote}
{\bf Figure 1:} {\small
The root diagram of $SO(5)$. With the Higgs vector $\bf h$ oriented as
in (a) the gauge symmetry is broken to $U(1)\times U(1)$, while with
the orientation in (b) the breaking is to $SU(2) \times U(1)$.}
\end{quote}
\vskip 0.40cm

\noindent of Sec.~6.  If instead $\bf h$ is
perpendicular to $\bgamma$, as in Fig.~1b, the unbroken gauge group is
$SU(2)\times U(1)$, with the roots of the $SU(2)$ being $\pm \bgamma$.
These are both orthogonal to $\bf g$, so Eq.~(\ref{purelyabelian}) is
satisfied and Eq.~(\ref{nonAtwoI}) tells us that there is again an
eight-parameter family of solutions.  It turns out that these
solutions, which are spherically symmetric, can be found 
explicitly \cite{so5}.  This makes it possible to determine
the background gauge zero modes and then use Eq.~(\ref{metricdef}) to
obtain the moduli space metric directly.  The result can then be
compared with the $m_\bgamma \rightarrow 0$ limit of the first case.

I will begin by describing the $SU(2)\times U(1)$ solutions.  Three of
its eight parameters give the location of the
center-of-mass.  Four others are phase angles that specify the
$SU(2)\times U(1)$ orientation.  (All elements of the unbroken group
act nontrivially on the solution.)  This leaves only a single
parameter, which I will denote by $b$, whose   
significance can
be found by examining the detailed form of the solutions.   To write these we
need some notation.  Let ${\bf t}(\balpha)$ and ${\bf t}(\bgamma)$ be defined as
in Eq.~(\ref{tripletdef}) and let
\begin{equation}
  {\cal M} = \frac{i }{ \sqrt{{\mbox{\boldmath $\beta$}}^2}} 
       \left(\matrix{E_{\mbox{\boldmath $\beta$}}  & 
      -E_{-{\mbox{\boldmath $\mu$}}} \cr
       E_{\mbox{\boldmath $\mu$}}  & \,\,\,
       E_{-{\mbox{\boldmath $\beta$}}}} \right)\, .
\end{equation}
Any adjoint representation $SO(5)$ field $P$ can then be decomposed into
parts that are respectively singlets, triplets, and doublets under the
unbroken $SU(2)$ by writing
\begin{equation}
   P = {\bf P}_{(1)} \cdot {\bf t}({\mbox{\boldmath $\alpha$}}) 
       + {\bf P}_{(2)} \cdot {\bf t}({\mbox{\boldmath $\gamma$}}) 
       + {\rm tr}\, P_{(3)} {\cal M} \, .
\end{equation}

With this notation, the solutions can be written as
\begin{eqnarray}
   A^a_{i(1)} &=& \epsilon_{aim}{\hat r_m} A(r)  \qquad\qquad \quad 
   \phi^a_{(1)}= {\hat r_a} H(r) \nonumber \\
   A^a_{i(2)} &=& \epsilon_{aim}{\hat r_m} G(r,b) \qquad\qquad
   \phi^a_{(2)} = {\hat r_a} G(r,b) \nonumber \\
   A_{i(3)} &=& \tau_i F(r,b) \qquad\qquad \qquad \,\,
   \phi_{(3)}= -i I F(r,b)  
\label{myansatz}
\end{eqnarray}
where $A(r)$ and $H(r)$ are the same as the coefficient functions in the
$SU(2)$ BPS monopole solution given in
Eq.~(\ref{SUtwoBPS}) and
\begin{eqnarray}
   F(r,b) &=& { v \over \sqrt{8} \cosh (vr/2) }  L(r, b)^{1/2} \nonumber \\
   G(r,b) &=& A(r) L(r, b) \label{cloud}
\end{eqnarray}
with
\begin{equation}
   L(r, b) = \left[ 1 +  (r/b) \coth(vr/2) \right]^{-1} 
\end{equation}
and $v = {\bf h}\cdot \balpha$.

The parameter $b$, which has dimensions of length, can take on any
positive real value.  It only enters into the doublet and
triplet components of the fields, and then only through the function
$L(r,b)$.  While the doublet fields decrease exponentially fast
outside the monopole core, the triplet fields have long-range
components whose character is determined by $b$.  For $1/v 
\la r \la b$ these
fall as $1/r$, resulting in a Coulomb magnetic field appropriate to a
non-Abelian magnetic charge.  At larger distances, however, the vector
potential falls as $1/r^2$, implying a field strength falling as
$1/r^3$ and thus showing that the magnetic charge is purely Abelian.
Thus, one might view these solutions as being composed of a massive
monopole, with a core of radius $\sim 1/v$, surrounded by a
``non-Abelian cloud'' of radius $\sim b$ that cancels the non-Abelian
part of its charge. 

In order to obtain the moduli space metric from Eq.~(\ref{metricdef}), we
need the background gauge zero modes about these solutions.  An
infinitesimal variation with respect to $b$ gives one zero mode, which
turns out to already be in background gauge.  The three $SU(2)$ modes
can then be obtained from this by a transformation of the type shown
in Eq.~(\ref{rightmult}).  The translational and $U(1)$ modes could also be
obtained in the usual fashion.  However, we do not need to do so,
since the corresponding parts of the metric can be inferred from the
BPS mass formulas for monopoles and dyons.  The result of all this is
\begin{equation}
    ds_{SU(2)\times U(1)}^2 = M d{\bf x}^2 +  {16\pi^2 \over M }d\chi^2 + 
       k \left[{ db^2 \over b} 
      + b \left( d\alpha^2 + \sin^2\alpha d\beta^2  
        + (d \gamma + \cos\alpha d\beta)^2 \right) \right]  
\label{NUSmetric}
\end{equation}
where $M$ is the monopole mass, $\bf x$ is the location of the
center of the monopole, $\chi$ is the $U(1)$ phase, and $\alpha$, $\beta$, and
$\gamma$ are the three angles specifying the $SU(2)$ orientation of
the solution.  The coefficient $k$ is a constant whose value is
unimportant for our purposes.

This should be compared with the two-monopole moduli space metric when
the symmetry is broken to $U(1)\times U(1)$.  Let $M=m_\bbeta
+m_\bgamma$ and $\mu = m_\bbeta m_\bgamma/M$ denote the total mass and
reduced mass of the system.  After transformation into
center-of-mass and relative variables, the metric given by
Eq.~(\ref{metric}) takes the form
\begin{eqnarray}
    ds_{U(1)\times U(1)}^2 &=& M d{\bf x}_{\rm cm}^2 +  {16\pi^2 \over M
    }d\chi_{\rm 
    tot}^2 +  \left(\mu +{k\over r}\right) \left[ dr^2 +r^2(d\theta^2
    + \sin^2\theta d\phi^2) \right]  \cr & & \qquad 
  + k^2 \left(\mu +{k\over r}\right)^{-1} (d\psi +d\cos\theta d\phi)^2 \, .
\end{eqnarray}
Here ${\bf x}_{\rm cm}$ specifies the position of the center-of-mass,
$r$, $\theta$ and $\phi$ are the spherical coordinates specifying the
relative positions of the two monopoles, $\chi_{\rm tot}$ and $\psi$
are overall and relative $U(1)$ phases, and $k$ is the same constant
as in Eq.~(\ref{NUSmetric}).  We are interested in the limit where
$m_\bgamma \rightarrow 0$ with $M$ held fixed.  In this limit the
reduced mass $\mu$ vanishes, and the metric becomes
\begin{equation}
     ds_{U(1)\times U(1)}^2 =  M d{\bf x}_{\rm cm}^2 +  {16\pi^2 \over M
          }d\chi_{\rm  tot}^2 
         +  k \left[{ dr^2 \over r} 
      + r \left( d\theta^2 + \sin^2\theta d\phi^2  
        + (d\psi + \cos\theta d\phi)^2 \right) \right] \, .
\end{equation}
This is exactly the same metric as in Eq.~(\ref{NUSmetric}), but with
a different notation: $b$ replaced by $r$, and $\alpha$, $\beta$, and
$\gamma$ replaced by $\theta$, $\phi$, and $\psi$, respectively.
Thus, the moduli space metric behaves smoothly in the limit where the
unbroken symmetry becomes non-Abelian, with the number of degrees of
freedom being conserved.  However, the interpretation of these
coordinates undergoes a curious change.  In particular, as one of the
monopoles becomes massless, its position becomes somewhat ambiguous.
While the separation $r$ goes over into the cloud radius $b$, which
has a definite gauge-invariant meaning, the directional angles
$\theta$ and $\phi$ are replaced by two global $SU(2)$ gauge phases.
Hence, two solutions with the same intermonopole separation but
different values for $\theta$ and $\phi$ are
physically distinct as long as the $\bgamma$-monopole remains massive,
but become gauge-equivalent when $m_\bgamma =0$.

\section{More complex examples}

   Further insight into the nature of the massless monopoles and the
non-Abelian cloud can be obtained by considering some more complex
solutions that arise in $SU(N)$ gauge theories.  The asymptotic value
of the adjoint Higgs field in some fixed direction can be brought into
the form
\begin{equation}
   \Phi_0 = {\rm diag}\,\, (t_N, t_{N-1}, \dots, t_1)
\label{phievalues}
\end{equation}  
with $t_1 \le t_2 \le \dots \le t_N$.  The set of simple roots picked
out by Eq.~(\ref{simplechoice}) then generate the $SU(2)$ subgroups
that lie in $2\times 2$ blocks along the diagonal and the magnetic
charge is given by 
\begin{equation}
   {\bf g}\cdot {\bf H} ={4\pi}\, {\rm diag}\,\, (n_{N-1},
           n_{N-2}-n_{N-1}, \dots,   n_1-n_2, -n_1 )  \, .
\label{Qevalues} 
\end{equation} 
If the $t_j$ are all unequal, the symmetry breaking is maximal, to
$U(1)^{N-1}$, and the $n_j$ are the topological charges.  Here I will
be primarily interested instead in the case where the middle $N-2$
eigenvalues of $\Phi_0$ are equal and the unbroken group is
$U(1)\times SU(N-2)\times U(1)$.  As explained previously, I will
focus on configurations in which the asymptotic
magnetic field is purely Abelian and commutes with all elements of the
unbroken $SU(N-2)$; i.e., configurations for which the middle $N-2$
eigenvalues of ${\bf g}\cdot {\bf H}$ are all equal.   

    All choices for the $\{n_j\}$ that satisfy this condition can be
written as combinations of three irreducible solutions\footnote{The
existence three types of irreducible solutions can be understood by
noting that the states corresponding to the two species of massive
monopoles in this theory transform under the $(N-2)$-dimensional fundamental
and antifundamental representations of $SU(N-2)$.  An
$SU(N-2)$ singlet can be formed from $N-2$ fundamentals, $N-2$
antifundamentals, or from a fundamental and an antifundamental.}:

\noindent 1)  $n_j = j-1$, so that 
\begin{equation}
   {\bf g}\cdot {\bf H} ={4\pi }\, {\rm diag}\,\, ((N-2),
           -1, -1, \dots, -1 , 0 )  \, .
\end{equation}
\noindent 2) $n_j = N-j-1$, so that 
\begin{equation}
   {\bf g}\cdot {\bf H} ={4\pi } \,{\rm diag}\,\, (0,
           1,1, \dots,  1 , -(N-2))  \, .
\end{equation}
\noindent 3) $n_j=1$ for all $j$, leading to  
\begin{equation}
   {\bf g}\cdot {\bf H} ={4\pi } \,{\rm diag}\,\, (1,
           0, 0, \dots,  0 , -1 )  \, .
\end{equation}
(Note that the moduli space metric for this case can be obtained from
the results of Sec.~6; the metrics for the first two cases are not
known.) 

Configurations of the first type can be viewed as containing $N-2$
massive and $(N-2)(N-3)/2$ massless monopoles, with the massive
monopoles all corresponding to the last simple root.  Eq.~(\ref{nonAtwoI})
shows that they depend on $2(N-1)(N-2)$ parameters.  Of these,
$4(N-2)$ presumably specify the positions and $U(1)$ phases of the
massive monopoles.  Specifying the $SU(N-2)$ orientation of the
configuration requires another ${\rm dim}\, [SU(N-2)]= (N-2)^2 -1$
parameters.  Hence, the remaining $(N-3)^2$ parameters describe
gauge-invariant aspects of the non-Abelian cloud, showing that it is
possible for this cloud to have considerably more structure than it
did in the $SO(5)$ example of the previous section.

Configurations of the second type also contain $(N-2)(N-3)/2$ massless
and $N-2$ massive monopoles, but now with the latter corresponding to
the first simple root.

Finally, configurations of the third type contain two massive
monopoles (one of each massive species), together with $N-3$ massless
monopoles.  There are $4(N-1)$ parameters in all, 8 of which specify
the positions and $U(1)$ phases of the massive monopoles.  One might
have expected to find an additional $(N-2)^2 -1$ parameters associated
with the unbroken $SU(N-2)$, as in the previous cases.  Except for
the simplest nontrivial case, $SU(4)$ broken to $U(1)\times SU(2)
\times U(1)$, there are clearly not enough parameters.  The
explanation is that, as we will see more explicitly below, any
configuration of this type for gauge group $SU(N)$ with $N>4$ can be
obtained by an embedding of an $SU(4)$ solution.  As a result, there 
are only ${\rm dim}\, [SU(N-2)/U(N-4)] = 4N-13$ global gauge
parameters.  There is but a single remaining parameter, which is
associated with the non-Abelian cloud.

Having explicit expressions for the solutions in these cases would
clearly be quite helpful for understanding the nature and
characteristics of the non-Abelian cloud.  Such expressions are not
known for the first two cases.  However, solutions for the third case
can be obtained explicitly, as I will now describe, by making use of
Nahm's construction of the BPS monopole solutions \cite{nahmpaper}.

The fundamental elements in Nahm's approach \cite{nahm} are a triplet
of matrices $T_a(t)$ that satisfy a set of nonlinear ordinary
differential equations.  These then define a set of linear
differential equations for another set of matrices, $v(t,{\bf r})$,
from which the spacetime fields ${\bf A}({\bf r})$ and ${\bf
\Phi}({\bf r})$ can be readily obtained.  I will now describe the
details of this construction for the case of a gauge group $SU(N)$.

The eigenvalues $t_j$ of $\Phi_0$ divide the range $t_1 \le t \le t_N$
into $N-1$ intervals.  On the $j$th interval, $t_j < t < t_{j+1}$, let
$k(t) \equiv n_j$, where $n_j$ is given by Eq.~(\ref{Qevalues}).  The matrices
$T_a(t)$ are required to have dimension $k(t) \times k(t)$.  In
addition, whenever two adjacent intervals have the same value for
$k(t)$, there are three matrices $\balpha_j$, of dimension $k(t_j)
\times k(t_j)$, defined at the interval boundary $t_j$.  These
matrices are required to obey the Nahm equation, 
\begin{equation} 
   {dT_a \over dt} = {i\over 2}\epsilon_{abc}[T_b,T_c] + \sum_j (\alpha_j)_a
             \delta(t-t_j) \, .
\label{Nahmeq}
\end{equation}
where the sum in the last term is understood to only run over those
values of $j$ for which the $\balpha_j$ are defined.  Having solved
this equation, one must next find a $2k(t) \times N$
matrix function $v(t,{\bf r})$ and $N$-component row vectors $S_j({\bf r})$
obeying the linear equation 
\begin{equation}
   0= \left[ -{d\over dt} + ({ T_a} + {r_a})\otimes { \sigma_a} \right]v 
  + \sum_j a_j^\dagger S_j \delta(t-t_j)
\label{veq}
\end{equation}
together with the orthogonality condition
\begin{equation} 
  I = \int dt\,  v^\dagger  v    + \sum_j S^\dagger_j S_j  \, .
\label{normalization}
\end{equation}
In Eq.~(\ref{veq}), $a_j$ is a $2k(t_j)$-component row vector obeying 
\begin{equation} 
     a^\dagger_j  a_j   = {\balpha}_j\cdot { \bsigma} -i (\alpha_j)_0 I
\label{adef}
\end{equation}
with $(\alpha_j)_0$  chosen so that the above matrix has rank 1.
Finally, spacetime fields obeying the BPS equations are given by
\begin{equation}
  \Phi =  \int dt \,t \,v^\dagger  v  + \sum_j t_j
            S^\dagger_j S_j 
\label{phieq}
\end{equation} 
\begin{equation}
  {\bf A} = -{i\over 2} \int dt \left[v^\dagger  {\bf \nabla} v 
           - {\bf \nabla} v^\dagger \,  v  \right]
         -{i\over 2}  \sum_j \left[ S^\dagger_j {\bf \nabla} S_j 
           -{\bf \nabla} S^\dagger_j S_j \right] \, .
\label{Aieq}
\end{equation} 

I will consider the case where the $n_j$ are all equal to
unity, so $k(t)=1$ over the entire range and there are an $\balpha_j$
and an $S_j$ for each value of $j$ from 2 through $N-1$.  To begin, I
will assume that the $t_j$ are all different, so that there are $N-1$
distinct massive monopoles, although I will soon
turn to the case with unbroken $U(1)\times SU(N-2)\times U(1)$ symmetry.
Eq.~(\ref{Nahmeq})
is solved by the piecewise constant solution 
\begin{equation}
   {\bf T}(t) = - {\bf x}_j     \, , \qquad  t_j < t < t_{j+1}  \, ,
\end{equation}
where the ${\bf x}_a$ have a natural interpretation as the positions of the
individual monopoles.  The $a_j$ of Eq.~(\ref{adef}) are
simply two-component row vectors that may be taken to be
\begin{equation} 
     a_j = \sqrt{2|{\bf x}_j-{\bf x}_{j-1}|} \left( \cos(\theta/2)
                   e^{-i\phi/2},  
                 \sin(\theta/2) e^{i\phi/2}\right)
\end{equation}
where $\theta$ and $\phi$ specify the direction of the vector
$\balpha_j={\bf x}_{j-1} -{\bf x}_j$.

Next, we must find a $2\times N$ matrix $v(t)$ and a set of
$N$-component row vectors $S_k$ ($k=2,3, \dots, N-1$) that satisfy
Eq.~(\ref{veq}).  To do this, let us first define a function $f_k(t)$
for each interval $t_k \le t \le t_{k+1}$, with
\begin{eqnarray}
     f_1(t) &=& e^{(t-t_2)({\bf r} - {\bf x}_1)\cdot { \bsigma}}  \cr   
     f_k(t) &=& e^{(t-t_k)({\bf r} - {\bf x}_k)\cdot { \bsigma}}    
         f_{k-1}(t_k)   \, , \qquad k >1  \, .
\label{fdef}
\end{eqnarray}
These are defined so that at the boundaries between intervals 
$f_k(t_k) = f_{k-1}(t_k)$.  An arbitrary solution of Eq.~(\ref{veq})
can then be written as
\begin{equation}
     v^a(t) = f_k(t) \eta_k^a \, , \qquad  t_k < t < t_{k+1} \, ,
\label{etadef}
\end{equation}
where the $\eta_k$ ($1 \le k \le N-1$) are a set of $N$-component
row vectors.   The discontinuities at the interval boundaries must be
such that 
\begin{equation}
      \eta_k=\eta_{k-1} + [f_k(t_k)]^{-1}_k a_k^\dagger S_k  \, .
\label{discontinuity}
\end{equation}
The normalization condition  Eq.~(\ref{normalization}), takes the form
\begin{equation} 
    I = \sum_{j = 2}^{N-1} 
            S_j ^\dagger S_j 
     + \sum_{k=1}^{N-1}  {\eta_k}^\dagger N_k \eta_k  
\label{orthog}
\end{equation}
with 
\begin{equation}
     N_k = \int_{t_k}^{t_{k+1}}dt \,f_k^\dagger(t) f_k(t)  \, .
\label{Ndef}
\end{equation}

These equations do not completely determine the $\eta_k$.  This
indeterminacy reflects the fact that Eq.~(\ref{veq}) is
preserved if $v$ and the $S_k$ are multiplied on the right by any
$N\times N$ unitary matrix function of $\bf r$; this corresponds to an
ordinary spacetime gauge transformation.  A convenient choice is to
take two columns of $v$, say $v^1$ and $v^2$, to be continuous.  This can be
done by setting $S_k^a =0$ for $a= 1, 2$ and choosing
\begin{equation}
    \eta_k^a = N^{-1/2} \theta^a  \, ,   \qquad a=1,2 \, ,
\end{equation}
with $N=\sum_k N_k$ and the $\theta^a$ being the two-component objects
$\theta^1=(1,0)^t$ and $\theta^2=(0,1)^t$.  Orthogonality of the other
columns of $v$ with the first two, as required by Eq.~(\ref{normalization}),
then implies that 
\begin{equation}
     0 = \sum_{k=1}^{N-1} N_k \eta^\mu_k 
\end{equation} 
where here and below Greek indices are assumed to run from 3 to $N$.
Together with the discontinuity Eq.~(\ref{discontinuity}), this uniquely
determines the $\eta_k^\mu$.  Substituting the result back into
Eq.~(\ref{orthog}) then gives an equation for the $S_k^\mu$, 
\begin{equation}
    \delta^{\mu\nu} = \sum_{i,j= 2}^{N-1} 
            {S_i^\mu}^\dagger [ \delta_{ij} + a_i M_{ij}a_j^\dagger ]
             S^\nu_j 
\label{Scondition}
\end{equation}
where the $M_{ij}$ are matrices, constructed from the $N_k$ and the
$f_k(t_k)$, whose precise form is not important for our purpose.
After solving this equation for the $S_k^\mu$, one can then work back to obtain
the $\eta_k^\mu$ and thus $v$, and then substitute into Eqs.~(\ref{phieq})
and (\ref{Aieq}) to obtain the spacetime fields.

Now let us specialize to the case of unbroken $U(1)\times SU(2)\times
U(1)$ symmetry.  The middle $N-2$ eigenvalues of $\Phi_0$ are now
degenerate, and so all but the first and last intervals in $t$
vanish.  Because the $f_k(t_k)$ are all equal to unity, the
discontinuity equation for the $\eta_k$ becomes
\begin{equation}
      \eta_k=\eta_{k-1} + a_k^\dagger S_k  \, .
\label{newdisc}
\end{equation}
In addition, the matrices $M_{ij}$ in Eq.~(\ref{Scondition}) no longer
depend on $i$ and $j$, but instead are all equal to a single matrix
$M$.  

When the symmetry breaking was maximal, the monopole positions entered
both through the functions $f_k(t)$ and through the $a_j$.  Now
however, with the middle intervals having zero width, the positions
associated with the massless monopoles enter only through the $a_j$.
But these now appear in Eqs.~(\ref{Scondition}) and (\ref{newdisc})
only in the combination $\sum_j a_j^\dagger S_j^\mu $.  This fact has
a striking consequence.  Consider two sets of monopole positions 
${\bf x}_k$ and $\tilde{\bf x}_k$ with identical positions for the 
massive monopoles, but with the massless
monopoles constrained only by the requirement that $\tilde a_j =W_{jk}
a_k$, where $W$ is any $(N-2)\times (N-2)$ unitary matrix.  If
$S_j^\mu$ is a solution of  Eq.~(\ref{Scondition}) for the first set
of positions, then $\tilde S_j^\mu = W_{jk} S_k^\mu$ is a
solution for the transformed set. 

This implies that the positions of the massless monopoles are not all
physically meaningful quantities.  This result was anticipated by the
parameter counting done earlier in this section, which indicated that
there should be a single gauge-invariant quantity characterizing the
non-Abelian cloud.  This quantity can be identified by noting that
these transformations leave invariant 
\begin{equation}
    \sum_j  a^\dagger_j a_j = 
 \sum_j
 \left[ {\bf\alpha}_j\cdot { \bsigma} -i {\alpha_j}_0 I \right]
     = ({\bf x}_1 - {\bf x}_{N-1})\cdot \bsigma 
         + \sum_{j=2}^{N-1} |{\bf x}_j - {\bf x}_{j-1}| \, .
\end{equation} 
The first term on the right hand side is determined by the positions
of the massive monopoles, while the second is just the sum of
the distances between successive massless monopoles.  The latter can be used
to define a cloud parameter $b$ by
\begin{equation}
    2b +R = \sum_{j=2}^{N-1} |{\bf x}_j - {\bf x}_{j-1}|
\end{equation}
where $R$ is the distance between the massive monopoles.

The subsequent analysis can be simplified by using a transformation of this
type to choose a canonical set of massless monopole positions in which
${\bf x}_2$ is located on the straight line defined by ${\bf x}_1$ and
${\bf x}_{N-1}$ at a distance $b$ from ${\bf x}_1$, while the
remaining $N-3$ massless monopoles are located at ${\bf x}_{N-1}$.
Once this choice is made, one is rather naturally led to choose a
solution for the $S_k$, and hence for the $\eta_k$ and $v$, such that
the resulting expressions for the spacetime fields have nontrivial
components only in a $4 \times 4$ block.  Thus, as promised earlier,
the solutions can all be obtained by embeddings of $SU(4) \rightarrow
U(1)\times SU(2) \times U(1)$ solutions.  

The fact that the solutions for arbitrary $SU(N)$ can be obtained from
the $SU(4)$ solution underscores the difficulties in pinning down the
massless monopoles.  When viewed as an $SU(4)$ solution, the
configuration contains a single massless monopole, but when it is
interpreted as an $SU(N)$ solution there are $N-3$ massless monopoles.
Thus, not only the positions, but even the number of massless
components is ambiguous.

These $SU(4)$ solutions have some features that are reminiscent of the
$SO(5)$ solutions discussed in the previous section.   The fields can
be decomposed into pieces that transform as triplets, doublets, and
singlets under the unbroken $SU(2)$.  Only the first two depend on
$b$, and then only through a single function $L$, which is now a
$2\times 2$ matrix.  Also, the triplet and doublet components of the
Higgs field are given in terms of the same spacetime functions as the
corresponding gauge field components, just as was the case with the
$SO(5)$ solution.

\def\bhyl{\hat{\bf y}_L}
\def\bhyr{\hat{\bf y}_R}

The detailed form of these solutions \cite{nahmpaper} is rather complex. 
However, some insight into the nature of the non-Abelian cloud can be obtained
by examining the asymptotic behavior of the fields well outside the
cores of the massive monopoles.  Consider first the case $b \gg R$.
If the distances $y_L$ and $y_R$ from a point $\bf r$ to the two massive
monopoles are both much less than $b$, the Higgs field and magnetic
field can be written in the form 
\begin{equation}
    \Phi({\bf r}) = U_1^{-1}({\bf r})
        \left( \matrix{  t_4  - {1\over 2y_R} & 0 & 0 & 0 \cr\cr
             0 & t_2+{1\over 2y_R} & 0 & 0 \cr\cr
             0 & 0 & t_2 -{1\over 2y_L} & 0 \cr\cr
             0 & 0 & 0 & t_1 +{1\over 2y_L} } \right) 
        U_1({\bf r})    + \cdots
\end{equation}
\begin{equation}
    {\bf B}({\bf r}) = U_1^{-1}({\bf r})
        \left( \matrix{ {\bhyr\over 2y_R^2} & 0 & 0 & 0 \cr\cr
             0 &  -{\bhyr\over 2y_R^2} & 0 & 0 \cr\cr
             0 & 0 &{\bhyl\over 2y_L^2} & 0 \cr\cr
             0 & 0 & 0 &  -{\bhyl\over 2y_L^2}   } \right) 
        U_1({\bf r}) + \cdots
\end{equation}
where $U_1({\bf r})$ is an element of $SU(4)$ and the dots represent
terms that are suppressed by powers of $R/b$, $y_L/b$, or $y_R/b$.
These are the fields that one would expect for two massive monopoles,
each of whose magnetic charges has both a $U(1)$ component and a
component in the unbroken $SU(2)$ that corresponds to the middle $2
\times 2$ block.  If instead $y \equiv (y_L +y_R)/2 \gg b$,
\begin{equation}
    \Phi({\bf r}) = U_2^{-1}({\bf r})
        \left( \matrix{ t_4 -{1\over 2y} & 0 & 0 & 0 \cr\cr
             0 & t_2 & 0 & 0 \cr\cr
             0 & 0 & t_2  & 0 \cr\cr
             0 & 0 & 0 & t_1 +{1\over 2y}  } \right) 
        U_2({\bf r})    + O(b/y^2)
\end{equation}
\begin{equation}
    {\bf B}({\bf r}) = U_2^{-1}({\bf r})
        \left( \matrix{  {\hat{\bf y}\over 2y^2} & 0 & 0 & 0 \cr\cr
             0 & 0 & 0 & 0 \cr\cr
             0 & 0 & 0 & 0 \cr\cr
             0 & 0 & 0 &  -{\hat{\bf y}\over 2y^2}} \right) 
        U_2({\bf r}) +  O(b/y^3) \, .
\end{equation}
Thus, at distances large compared to $b$ the non-Abelian part of the
Coulomb magnetic field is cancelled by the cloud, in a manner similar
to that which we saw for the $SO(5)$ case.   

In the opposite limit, $b=0$, the solutions are essentially embeddings
of $SU(3) \rightarrow U(1) \times U(1)$ solutions.  At large
distances, one finds that 
\begin{equation}
    \Phi({\bf r}) = U_3^{-1}({\bf r})
        \left( \matrix{ t_4 -{1\over 2y_R} & 0 & 0 & 0 \cr\cr
             0 & t_2 -{1\over 2y_L} + {1\over 2y_R} & 0 & 0 \cr\cr
             0 & 0 & t_2 & 0 \cr\cr
             0 & 0 & 0 &  t_1 +{1\over 2y_L}} \right) 
        U_3({\bf r})
\end{equation}
\begin{equation}
    {\bf B}({\bf r}) = U_3^{-1}({\bf r})
        \left( \matrix{  {\bhyr\over 2y_R^2} & 0 & 0 & 0 \cr\cr
             0 & {\bhyl\over 2y_L^2} - {\bhyr\over 2y_R^2} & 0 & 0 \cr\cr
             0 & 0 & 0 & 0 \cr\cr
             0 & 0 & 0 &   -{\bhyl\over 2y_L^2}    } \right) 
        U_3({\bf r})  \, .
\label{b0asymB}
\end{equation}
Viewed as $SU(3)$ solutions, the long-range fields are purely
Abelian.  Viewed as $SU(4)$ solutions, the long-range part is
non-Abelian in the sense that the unbroken $SU(2)$ acts nontrivially
on the fields.  However, because of the alignment of the fields of the
two massive monopoles, the non-Abelian part of the field is a purely
dipole field that falls as $R/y^3$ at large distances.  

\section{Concluding remarks}

I have shown in these talks how one is naturally led to a class of
multimonopole solutions that contain one or more massive monopoles,
similar to those found in the $SU(2)$ gauge theory, surrounded by a
cloud, of arbitrary size, in which there are nontrivial non-Abelian
fields.  Analysis of the moduli space Lagrangian that governs the
low-energy monopole dynamics suggests that these clouds can be
understood in term of the degrees of freedom of massless monopoles
carrying purely non-Abelian magnetic charges. 

There remain many open questions relating to these massless monopoles.
First, it would clearly be desirable to obtain additional solutions
containing non-Abelian clouds.  Particularly useful would be solutions
with charges such that the cloud depends on more than a single
gauge-invariant parameter, and solutions containing more than a single
cloud.  Experience with the solutions described in Sec.~9 suggests
that, as a first step, it might be feasible to attack the simplified
problem of determining the cloud structure for a given set of massive
monopole positions.  From the more physical viewpoint, one would like
to use these solutions to gain further insight in the properties of
non-Abelian gauge theories.  The massless monopoles clearly seem to be
the duals of the massless gauge bosons.  Hence, one should be able to
find some kind of correspondence between the behavior of the
non-Abelian clouds and that of the gauge bosons.  Understanding this
correspondence in detail remains an important challenge.

\vskip .4cm

This work was supported in part by the U.S. Department of Energy.

\end{document}